\documentclass[fleqn,usenatbib]{mnras}
\usepackage{newtxtext,newtxmath}

\usepackage[T1]{fontenc}

\DeclareRobustCommand{\VAN}[3]{#2}
\let\VANthebibliography\thebibliography
\def\thebibliography{\DeclareRobustCommand{\VAN}[3]{##3}\VANthebibliography}

\usepackage{graphicx}	
\usepackage{amsmath}	
\usepackage{comment}

\title[NN 3D Tomography]{DeepCHART: Mapping the 3D dark matter density field from Ly$\alpha$ forest surveys using deep learning}

\author[Maitra et al.]{
Soumak Maitra,$^{1,2}$\thanks{E-mail: soumak.maitra@theory.tifr.res.in}
Matteo Viel,$^{3,2,4,5}$
and Girish Kulkarni$^{1}$
\\
$^{1}$Tata Institute of Fundamental Research, Homi Bhabha Road, Mumbai 400005, India\\
$^{2}$Istituto Nazionale di Astrofisica -Osservatorio Astronomico di Trieste, Via Tiepolo 11, Trieste, Italy\\
$^{3}$SISSA International School for Advanced Studies, Via Bonomea 265, 34136, Trieste, Italy\\
$^{4}$INFN - Sezione di Trieste, via Valerio 2, 34127, Trieste, Italy\\
$^{5}$IFPU, Institute for Fundamental Physics of the Universe, Via Beirut 2, 34014 Trieste, Italy
}

\date{Accepted ---. Received ---; in original form ---}

\pubyear{\the\year{}}

\begin{document}
\label{firstpage}
\pagerange{\pageref{firstpage}--\pageref{lastpage}}
\maketitle

\begin{abstract}
We present \textsc{DeepCHART} (Deep learning for Cosmological Heterogeneity and Astrophysical Reconstruction via Tomography), a deep learning framework designed to reconstruct the three-dimensional dark matter density field at redshift $z = 2.5$ from Ly$\alpha$ forest spectra. Leveraging a 3D variational autoencoder with a U-Net architecture, \textsc{DeepCHART} performs fast, likelihood-free inference, accurately capturing the non-linear gravitational dynamics and baryonic processes embedded in cosmological hydrodynamical simulations. When applied to joint datasets combining Ly$\alpha$ forest absorption and coeval galaxy positions, the reconstruction quality improves further. For current surveys, such as Subaru/PFS, CLAMATO, and LATIS, with an average transverse sightline spacing of $d_\perp = 2.4h^{-1}\mathrm{cMpc}$, \textsc{DeepCHART} achieves high-fidelity reconstructions over the density range $0.4<\Delta_{\rm DM}<15$, with a voxel-wise Pearson correlation coefficient of $\rho \simeq 0.77$. These reconstructions are obtained using Ly$\alpha$ forest spectra with signal-to-noise ratios as low as 2 and instrumental resolution $R=2500$, matching Subaru/PFS specifications. For future high-density surveys enabled by instruments such as ELT/MOSAIC and WST/IFS with $d_\perp\simeq 1h^{-1}\mathrm{cMpc}$, the correlation improves to $\rho \simeq 0.90$ across a wider dynamic range ($0.25<\Delta_{\rm DM}<40$). The framework reliably recovers the dark matter density PDF as well as the power spectrum, with only mild suppression at intermediate scales. In terms of cosmic web classification, \textsc{DeepCHART} successfully identifies 81\% of voids, 75\% of sheets, 63\% of filaments, and 43\% of nodes. We propose \textsc{DeepCHART} as a powerful and scalable framework for field-level cosmological inference, readily generalisable to other observables, and offering a robust, efficient means of maximising the scientific return of upcoming spectroscopic surveys.

\end{abstract}




\section{Introduction}

\begin{figure*}
    \centering
    \includegraphics[width=16cm]{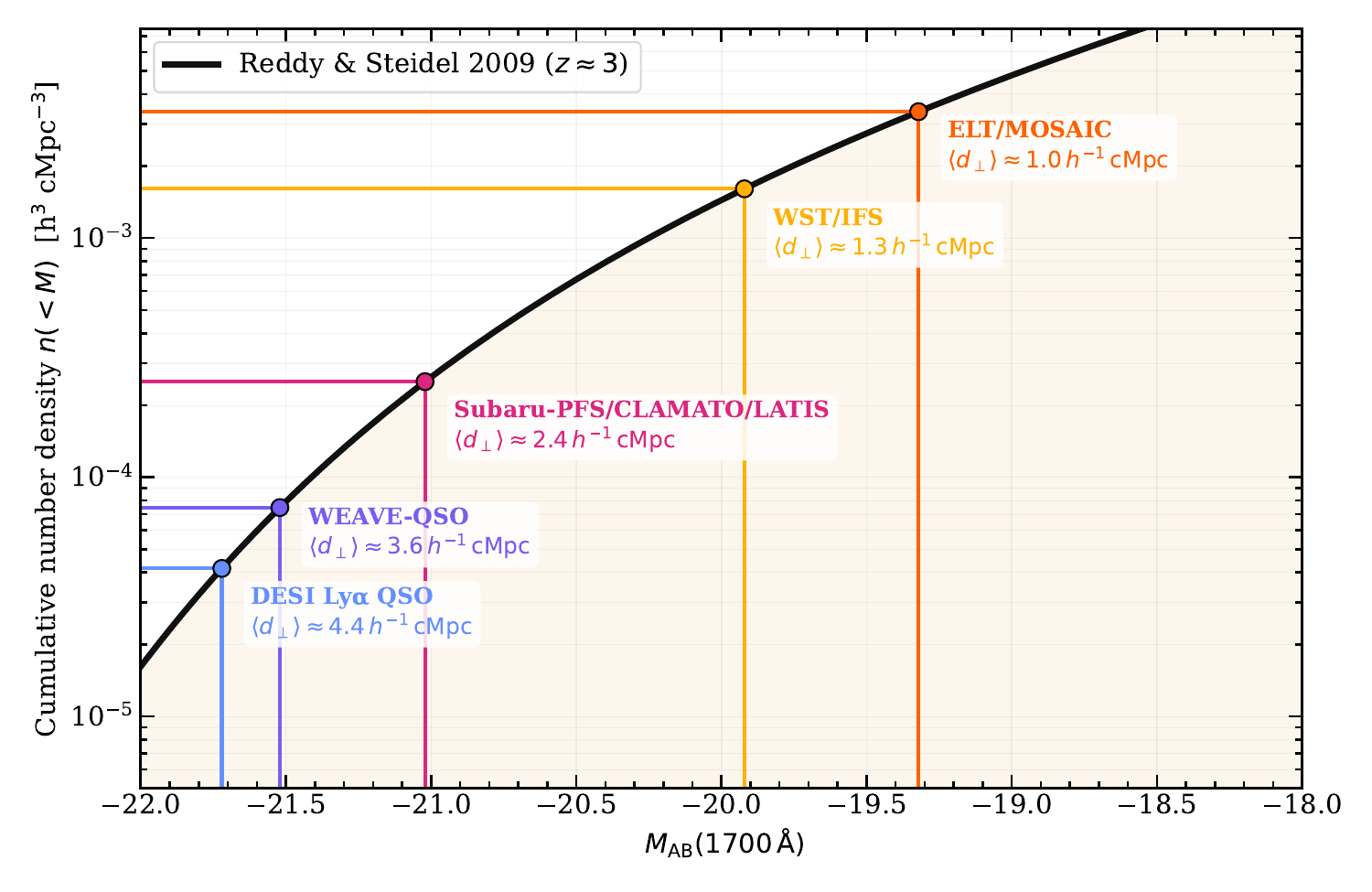}
    \caption{Cumulative number density of star-forming galaxies as a function of rest-frame UV absolute magnitude ($M_{\mathrm{AB}}(1700,\text{\AA})$) at $z \approx 3$, based on the luminosity function of \citet{Reddy2009}. Colored points and lines indicate the limiting magnitude, cumulative sightline density, and mean transverse separation ($d_\perp$) achievable by current and upcoming spectroscopic surveys targeting Ly$\alpha$ forest tomography. Labels highlight major projects, including DESI Ly$\alpha$ QSO, WEAVE-QSO, Subaru-PFS/CLAMATO/LATIS, WST/IFS, and ELT/MOSAIC. The plotted densities assume galaxies bright enough to yield Ly$\alpha$ forest spectra with signal-to-noise ratios as low as 2. Increasing survey depth enables smaller $d_\perp$, which is essential for high-fidelity tomographic mapping of the IGM on megaparsec scales.}
    \label{fig:survey}
\end{figure*}

The intergalactic medium (IGM) at high redshift constitutes the primary reservoir of baryonic matter in the Universe and traces the filamentary network of large-scale structure shaped by gravitational instability \citep{Meiksin2009, McQuinn2016}. On scales of a few megaparsecs and above, the spatial distribution of the IGM closely follows the underlying dark matter density field, making it a powerful probe of the cosmic web and the growth of structure \citep{Bi1997, Croft1998}. At smaller scales, particularly on $\sim1$Mpc and below, the correspondence between the IGM and dark matter becomes increasingly influenced by baryonic physics, including gas pressure smoothing, thermal broadening, and radiative feedback processes \citep{miralda1996,schaye2000, mcdonald2001, schaye2001, aguirre2001, viel2002, oppenheimer2006, Bolton2008, viel2013, Bolton2017}. These non-gravitational effects imprint complex signatures on the IGM, such that its small-scale structure reflects not only the underlying dark matter distribution but also the detailed thermal and ionization history of the Universe \citep{peeples2010a, peeples2010b,  rorai2017,rorai2018}. As a result, proper understanding and modeling of IGM observables on these scales require accounting for the intertwined influences of both dark matter and baryonic physics.

A powerful observational window into this diffuse cosmic structure is provided by the Ly$\alpha$ forest, a dense series of absorption features imprinted on the spectra of distant quasars and galaxies by intervening neutral hydrogen in the IGM \citep{Gunn1965, Lynds1971, Rauch1998}. The Ly$\alpha$ forest has traditionally been analyzed as a one-dimensional probe along isolated sightlines. Its clustering properties have been extensively studied using two-point statistics such as the flux power spectrum \citep{mcdonald2000, McDonald2006, seljak2006, zaroubi2006}, as well as higher-order statistics like the three-point correlation function or bispectrum, which capture non-Gaussian features \citep{viel2004a, maitra2019, maitra2022a, maitra2022b}. Over the past two decades, advances in spectroscopic surveys have enabled the tomographic reconstruction of three-dimensional IGM absorption maps using dense grids of background sources \citep{Pichon2001, Caucci2008, Lee2014, Horowitz2021}.
The fidelity of such reconstructions, especially on $\sim1$ Mpc scales and below, ultimately depends on accurately modeling the interplay between gravitational evolution and the complex baryonic physics that shape the IGM.

Early feasibility studies proposed that a sufficiently dense network of sightlines—on the order of hundreds per square degree—could accurately map the IGM and robustly identify cosmic structures such as filaments, clusters, and voids at redshifts $z \sim 2$--$3$ on megaparsec scales \citep{Pichon2001, Caucci2008, Lee2014}. Initial tomographic observations using faint galaxies as background sources in the COSMOS field helped demonstrate the practical feasibility of this approach \citep{Lee2014}. Subsequent large-scale observational campaigns, notably the COSMOS Ly$\alpha$ Mapping And Tomography Observations (CLAMATO), achieved sightline densities of approximately $150$ per square degree, enabling the reconstruction of volumes of $\sim 100h^{-3}\mathrm{Mpc}^3$ at resolutions of $\sim 2.5 h^{-1}$Mpc \citep{Lee2018}. These surveys have facilitated detailed studies of high-redshift cosmic voids, protoclusters, and the filamentary structure of the cosmic web \citep{Stark2015, Krolewski2018, Newman2020}.

Traditional tomographic reconstruction methods typically rely on analytical approximations and assumptions of Gaussianity, such as Wiener filtering or Gaussian smoothing, to interpolate the sparse 1D Ly$\alpha$ forest data into a continuous 3D density field \citep[e.g.][]{Lee2014, Lee2018}. Although computationally efficient, these approaches inherently smooth out small-scale, non-linear features of the cosmic web, limiting reconstruction fidelity at scales of a few megaparsecs and below \citep{Cisewski2014}. More advanced semi-analytical and iterative methods, notably the Tomographic Absorption Reconstruction and Density Inference Scheme (TARDIS), introduced constrained maximum-likelihood approaches that incorporate physically motivated priors based on nonlinear gravitational evolution and galaxy distributions \citep{Horowitz2019, Horowitz2021}. In TARDIS, the forward modeling of Ly$\alpha$ forest absorption is performed using the Fluctuating Gunn-Peterson Approximation (FGPA), which remains valid on scales of a few $h^{-1}$Mpc and larger, where the IGM density field closely traces the underlying dark matter distribution. While capable of improving spatial resolution and recovering small-scale cosmic structure, these methods are comparatively more computationally intensive, as they involve iterative forward modeling with \textsc{FastPM} \citep{feng2016} gravitational evolution for each reconstruction.

Recent and upcoming spectroscopic surveys are rapidly advancing the frontier of Ly$\alpha$ forest tomography by increasing the density of background sources, thereby reducing the mean transverse sightline separation and enhancing the fidelity of IGM mapping. As summarized in Figure~\ref{fig:survey}, current programs such as DESI Ly$\alpha$ QSO \citep{DESICollaboration2016, DESI2024} and WEAVE-QSO \citep{Pieri2016} achieve mean sightline separations of $d_\perp \approx 4.4h^{-1}$cMpc and $3.6h^{-1}$cMpc, respectively, through spectroscopy of bright quasars and galaxies at $z \approx 3$. Intermediate-depth surveys including Subaru-PFS, CLAMATO and LATIS \citep{Lee2014, Lee2018, Takada2014, Newman2020} reach densities corresponding to $d_\perp \approx 2.4h^{-1}$cMpc by targeting fainter star-forming galaxies. Looking ahead, future facilities will deliver a transformative leap in tomographic resolution. Next-generation facilities such as the Wide-field Spectroscopic Telescope (WST/IFS) and the Extremely Large Telescope (ELT) with the MOSAIC instrument \citep{Japelj2019,Mainieri2024} are poised to further extend these capabilities by combining large collecting areas with highly multiplexed spectroscopy.

It is important to note that the magnitude limits commonly reported for these surveys are typically based on continuum signal-to-noise thresholds of $\mathrm{S/N} \sim 3$--$4$, which sets a conservative floor for effective sightline density. If the S/N threshold is relaxed to $\sim 2$, the number of usable sightlines increases substantially, reducing the mean inter-sightline separation. As illustrated in Figure~\ref{fig:survey}, under these conditions, future facilities such as WST/IFS and ELT/MOSAIC could achieve mean separations as small as $d_\perp \approx 1.3h^{-1}$cMpc and $1.0h^{-1}$cMpc, respectively. This progression enables tomographic reconstructions of the IGM on scales approaching $1\,h^{-1}$cMpc, where the influence of complex baryonic processes, as resolved by hydrodynamical simulations, become increasingly significant. In this regime, it is therefore desirable to develop tomographic reconstruction methods that can incorporate the detailed physics captured in hydrodynamical models, thereby complementing and extending beyond traditional FGPA-based approaches for interpreting high-resolution maps of the cosmic web and the relationship between galaxies and their environments at high redshift.

Concurrently, cosmology has seen rapid advances from machine learning (ML) techniques, especially deep learning methods capable of learning complex mappings from high-dimensional data \citep{Ntampaka2019, VillaescusaNavarro2021}. Generative models, such as Variational Autoencoders (VAEs; \citealt{Kingma2013}), Generative Adversarial Networks (GANs; \citealt{Goodfellow2014}), and Diffusion models \citep{mishra2025}, have been successfully employed to reproduce non-linear cosmic structures and enable fast hydrodynamic field generation \citep{Lanusse2021, Jacobus2023, Rigo2025}. Notably, \citet{Horowitz2022} demonstrated that VAEs trained on cosmological simulations could accurately map dark matter distributions to baryonic fields, capturing complex nonlinear relationships beyond the reach of conventional analytical models. Complementing these generative approaches, discriminative methods such as information-maximizing neural networks \citep{charnock2018,maitra2024} have shown promise for directly constraining cosmological and astrophysical parameters from simulated observables, highlighting the versatility of deep learning across both inference and generation tasks.

Motivated by these developments, we introduce a novel deep learning approach to Ly$\alpha$ forest tomography, employing a 3D UNet-based Variational Autoencoder (VAE) architecture. Our method reconstructs the three-dimensional dark matter density field directly from sparse Ly$\alpha$ absorption sightlines and galaxy positions, leveraging training data from fully hydrodynamical cosmological simulations. While previous methods often rely on analytic approximations such as FGPA, which provides an effective description at relatively large scales and in the linear regime, the deep learning framework adopted here is able to naturally capture the small-scale nonlinearities and baryonic processes present in hydrodynamical simulations, particularly at scales approaching $1h^{-1}$cMpc and below.
Our deep learning framework further distinguishes itself by being inherently likelihood-free, eliminating the need for expensive iterative forward modeling during reconstruction. Once trained, the network can instantly generate statistically consistent 3D density fields from observational data, providing a probabilistic characterization of uncertainties through latent space sampling. This rapid inference capability is particularly valuable given the large data volumes anticipated from upcoming high-density spectroscopic surveys.

In this paper, we present a comprehensive overview of the \textsc{DeepCHART} framework, beginning in Section~\ref{sec:Simulation} with the construction of our training dataset from a suite of fully hydrodynamical simulations performed with \textsc{GADGET-3}, including realistic sightline sampling, spectral noise, and instrumental resolution. Section~\ref{sec:3dunet_vae} describes the architecture and training of our three-dimensional variational autoencoder (VAE) based on a U-Net backbone, outlining how sparse Ly$\alpha$ forest and galaxy inputs are mapped to full 3D dark matter density fields. In Section~\ref{sec:results}, we evaluate the reconstruction accuracy under current and future survey configurations, analysing voxel-level correlation statistics, density distribution, power spectrum recovery, and cosmic web classification performance across smoothing scales. We present visual comparisons of the reconstructed and true fields, and examine how tracer density and tracer type impact recovery. Section~\ref{sec:discussion} discusses and summarizes our key findings and outlines future directions.

Throughout this paper we assume a flat $\Lambda$CDM cosmology consistent with \cite{planck2016} measurements characterized by the parameters $(\Omega_{\Lambda}, \Omega_{m}, \Omega_{b}, h, n_s, \sigma_8, Y) = (0.678, 0.322, 0.0482, 0.678, 0.96, 0.829, 0.24)$. Comoving distances are quoted in units of $h^{-1}$cMpc. 

\begin{figure*}
    \centering
    \includegraphics[width=18cm, trim={0cm 1.5cm 0cm 1cm}, clip]{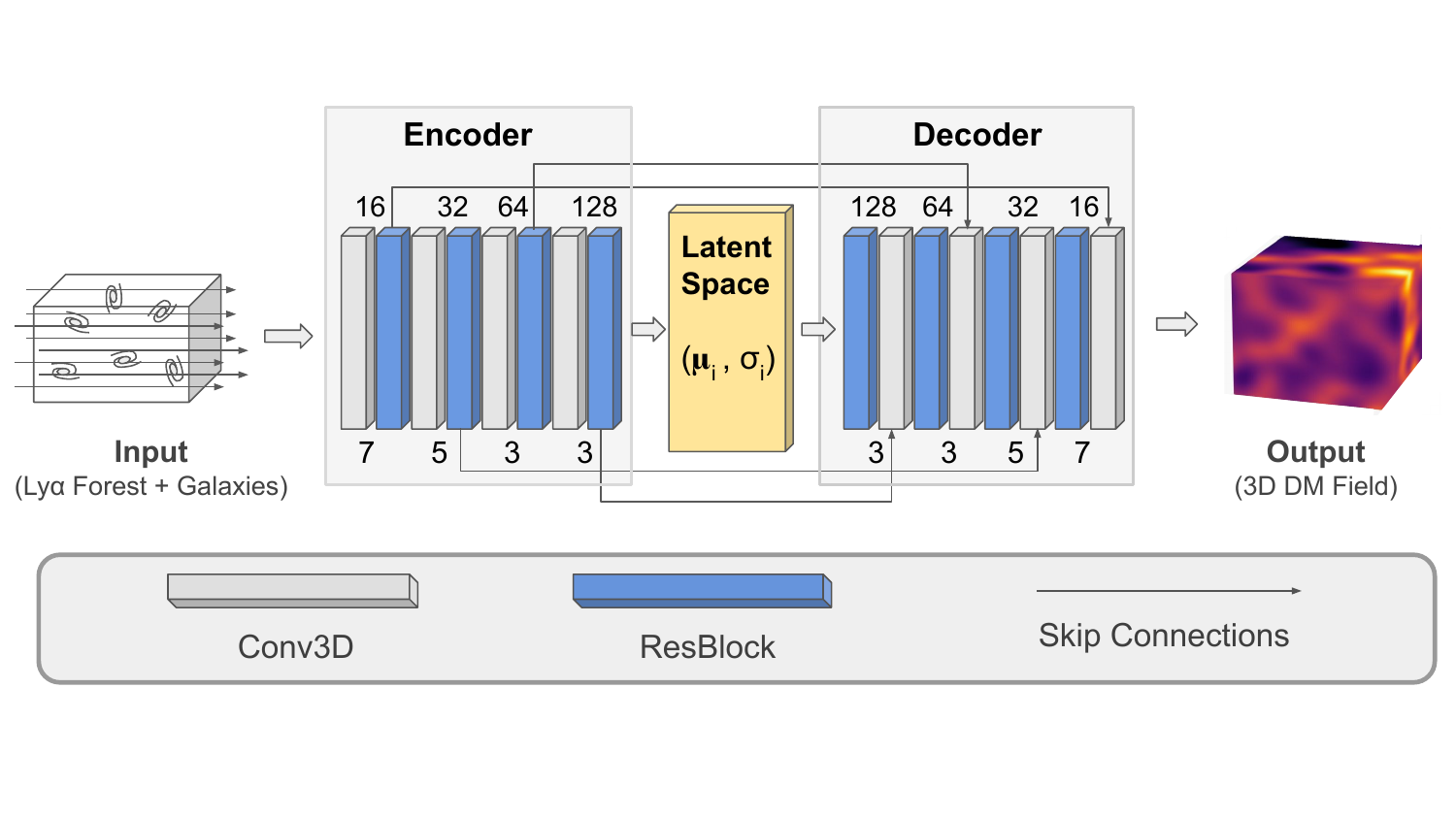}
    \caption{Schematic architecture of our 3D U-Net-based Variational Autoencoder (VAE) used for tomographic reconstruction of the dark matter density field. The network takes as input a stack of Ly$\alpha$ forest sightlines and galaxies, which are processed by the encoder through successive 3D convolutional layers and residual blocks, progressively compressing the data into a low-dimensional latent space parameterized by mean ($\mu_i$) and variance ($\sigma_i$) vectors. The latent space is 512-dimensional, allowing compact yet expressive encoding of the input volume. The decoder then reconstructs the three-dimensional dark matter (DM) field from the latent representation via a mirrored sequence of convolutional layers and residual blocks. Skip connections link corresponding encoder and decoder layers, facilitating efficient information flow and improved recovery of spatial structure. Kernel sizes are indicated below each layer, and the number of channels, indicating feature map dimensionalities, are indicated above. This architecture enables efficient learning of complex, non-linear mappings from sparse sightline data to full 3D density fields.}
    \label{fig:unetvae}
\end{figure*}
                                        
\section{Simulation} \label{sec:Simulation}

For our simulations, we utilize the smoothed particle hydrodynamics (SPH) code {\sc gadget-3} \citep[a modified version of the publicly available {\sc gadget-2}\footnote{\url{http://wwwmpa.mpa-garching.mpg.de/gadget/}}, see][]{springel2005}. This code incorporates radiative heating and cooling processes by self-consistently solving the coupled ionization and thermal evolution equations under a specified ultraviolet (UV) background. The simulations adopt the same physics of the state-of-the-art models based on the Sherwood-Relics suite of cosmological hydrodynamical simulations \cite{puchwein2023,Bolton2017}, which implies that the thermal history is fixed to match observational constraints.
Our simulations are performed within a periodic cubic box of size $(40 \, h^{-1}$ cMpc$)^3$, employing $2\times512^3$ particles. We adopt the standard flat $\Lambda$CDM cosmology characterized by the parameters $(\Omega_{\Lambda}, \Omega_{m}, \Omega_{b}, h, n_s, \sigma_8, Y) = (0.69, 0.31, 0.0486, 0.674, 0.96, 0.83, 0.24)$. The ionization and heating rates are derived from \citet{puchwein2019}. Initial conditions are set at redshift $z = 99$, generated via the Zeldovich approximation. The gravitational softening length is fixed at $1/30^{\mathrm{th}}$ of the mean inter-particle separation. To enhance computational efficiency, gas particles surpassing overdensity thresholds of $\Delta>10^3$ and temperatures below $10^5$ K are converted into collisionless star particles using the {\sc quick\_lyalpha} flag \citep[see][]{viel2004a}. Feedback mechanisms from active galactic nuclei (AGN), stellar processes, or galactic winds are not included. 

We have run a suite of 10 simulations with identical cosmological and astrophysical parameters but different initial random seeds. Nine of these are used to train the neural network for tomographic reconstruction, while the remaining one is reserved as an independent test set to evaluate reconstruction performance.

\subsection{\texorpdfstring{Generation of Line-of-Sight Spectra for Ly$\alpha$ Forest Tomography}{Generation of Line-of-Sight Spectra for Lya Forest Tomography}}

We simulate one-dimensional Ly$\alpha$ absorption spectra along sparse lines of sight through a cosmological volume, designed to replicate conditions expected for current and future Ly$\alpha$ tomography surveys. In particular, we consider two representative sightline separations: a mean transverse separation of $2.4\,h^{-1}\mathrm{cMpc}$, consistent with the projected sightline density of the Subaru Prime Focus Spectrograph (PFS) Deep Survey and similar programs \citep{Lee2014, Lee2018, Takada2014, Newman2020}, and a finer separation of $1.0\,h^{-1}\mathrm{cMpc}$, representative of the higher source densities anticipated for next-generation facilities \citep{Japelj2019,Mainieri2024}. For clarity, throughout this work we refer to the $2.4\,h^{-1}\mathrm{cMpc}$ case as the ``current'' survey scenario, and the $1.0\,h^{-1}\mathrm{cMpc}$ case as the ``future'' survey scenario. We assess tomographic reconstruction performance using mock sightlines generated for both these cases.

The sightlines are oriented along the $z$-axis and span the full comoving depth of the simulation box. We initially discretize each line of sight into 1024 evenly spaced grid points. At each grid point, physical quantities required for computing the Ly$\alpha$ optical depth---namely the neutral hydrogen number density $n_{\mathrm{HI}}$, gas temperature $T$, and line-of-sight peculiar velocity $v_{\mathrm{pec}}$---are computed using smoothed particle hydrodynamics (SPH) interpolation:
\begin{equation}
f_i = \sum_j f_j \frac{m_j}{\rho_j} W(|\mathbf{r}_i - \mathbf{r}_j|, h_j),
\end{equation}
where $f_j$, $m_j$, and $\rho_j$ are the quantity, mass, and density of SPH particle $j$, respectively, and $W$ is the cubic spline kernel:
\begin{equation}
W(r, h) = \frac{8}{\pi h^3}
\begin{cases}
1 - 6\left(\frac{r}{h}\right)^2 + 6\left(\frac{r}{h}\right)^3, & 0 \leq \frac{r}{h} \leq \frac{1}{2}, \\
2\left(1 - \frac{r}{h}\right)^2, & \frac{1}{2} < \frac{r}{h} \leq 1, \\
0, & \frac{r}{h} > 1.
\end{cases}
\end{equation}

Thermal Doppler broadening and line-of-sight peculiar velocity gradients are incorporated into the computation of the optical depth \(\tau\) using the standard Voigt profile formalism. The corresponding transmitted flux is then obtained via \(F = e^{-\tau}\). To match the spectral pixel size of observational data, the flux is regridded onto 384 evenly spaced points along the line of sight. We adopt the hydrogen ionization state directly from the simulation output without rescaling the photoionization rate, i.e., \(C_\Gamma = 1\). The simulations follow the ionization history of \citet{puchwein2019}, which yields a mean Ly\(\alpha\) forest flux consistent with observational measurements.

\subsection{Forward Modeling of Spectral Resolution and Noise Properties}

To compare these simulated spectra with real observational data, we forward model instrumental effects based on Subaru-PFS specifications. Each ideal transmitted flux spectrum is first convolved with a Gaussian kernel to emulate the instrumental line spread function (LSF). The full width at half maximum (FWHM) is taken as \( R = \lambda/\Delta\lambda = 2500 \), which corresponds to a velocity resolution of approximately \( \Delta v \sim 120\,\mathrm{km\,s^{-1}} \) at the Ly$\alpha$ wavelength \citep{PFSparams}:
\begin{equation}
\mathrm{FWHM} = \frac{c}{R} \approx \frac{3 \times 10^5\,\mathrm{km\,s^{-1}}}{2500} \approx 120\,\mathrm{km\,s^{-1}}.
\end{equation}

After Gaussian convolution, each spectrum is regridded from 1024 to 384 pixels to match the spectral resolution and binning of the final data products expected from the Subaru-PFS Deep Field survey. This binning corresponds to a pixel scale of roughly \( \sim 25\,\mathrm{km\,s^{-1}} \) per pixel.

To simulate realistic observational noise, we adopt a signal-to-noise ratio (SNR) distribution modeled after the TARDIS forward-modeling pipeline \citep{Horowitz2019}. The SNR per pixel is drawn from a probability distribution function approximated by a power-law \citep{Stark2015}:
\begin{equation}
P(\mathrm{SNR}) \propto \mathrm{SNR}^{-3.6}, \quad \text{for } 2 < \mathrm{SNR} < 10.
\end{equation}
This distribution reflects the flux-limited selection and exposure-time variations of quasars observed in the PFS Deep survey.

Gaussian noise is then added to each pixel according to the sampled SNR:
\begin{equation}
F_{\mathrm{noisy}}(x) = F_{\mathrm{conv}}(x) + \mathcal{N}(0, \sigma^2(x)), \ \ \text{where} \ \sigma(x) = \frac{F_{\mathrm{conv}}(x)}{\mathrm{SNR}(x)}.
\end{equation}
The resulting dataset provides a realistic mock catalog of Ly$\alpha$ forest sightlines suitable for tomographic reconstruction and cosmological inference. For consistency and to facilitate direct comparison, we apply this instrumental and noise modeling identically to both the current ($d_\perp \approx 2.4\,h^{-1}$cMpc) and future ($d_\perp \approx 1.0\,h^{-1}$cMpc) sightline separation scenarios, as the spectral resolution and typical exposure times of the surveys are expected to be comparable.

For each tomographic snapshot, we embed these one-dimensional spectra into a three-dimensional volume discretized into a grid of size \(128 \times 128 \times 384\), corresponding to the \((40\,h^{-1}\mathrm{cMpc})^3\) box. The sightlines occupy specific transverse positions in the \(x\)–\(y\) plane and are assigned flux values along the \(z\)-axis. Voxels not intersected by a sightline remain zero, resulting in a sparse 3D volume that mimics the observational geometry of Ly$\alpha$ forest tomography. This representation enables direct ingestion into the neural network pipeline and ensures consistency with the spatial anisotropy and pixel size of real spectral data.

\begin{figure*}
    \centering
    \includegraphics[width=4.9cm, trim={1cm 0cm 11.5cm 0cm}, clip]{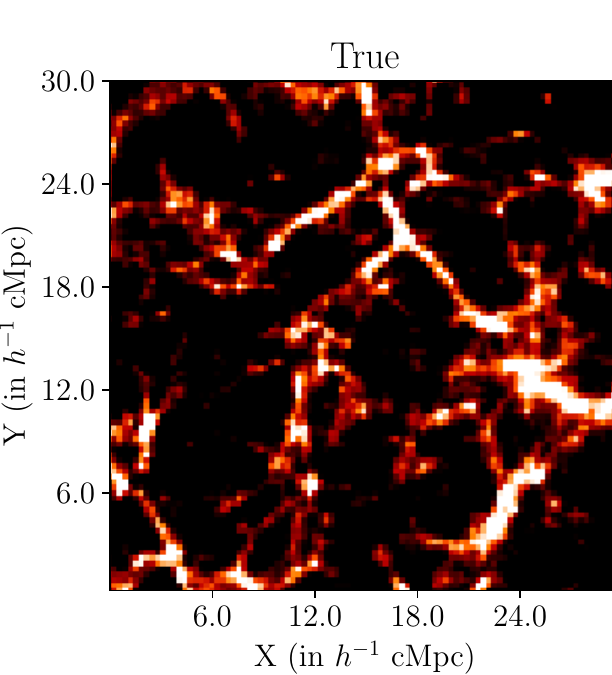}%
    \includegraphics[width=8.25cm, trim={2.6cm 0cm 2.8cm 0cm}, clip]{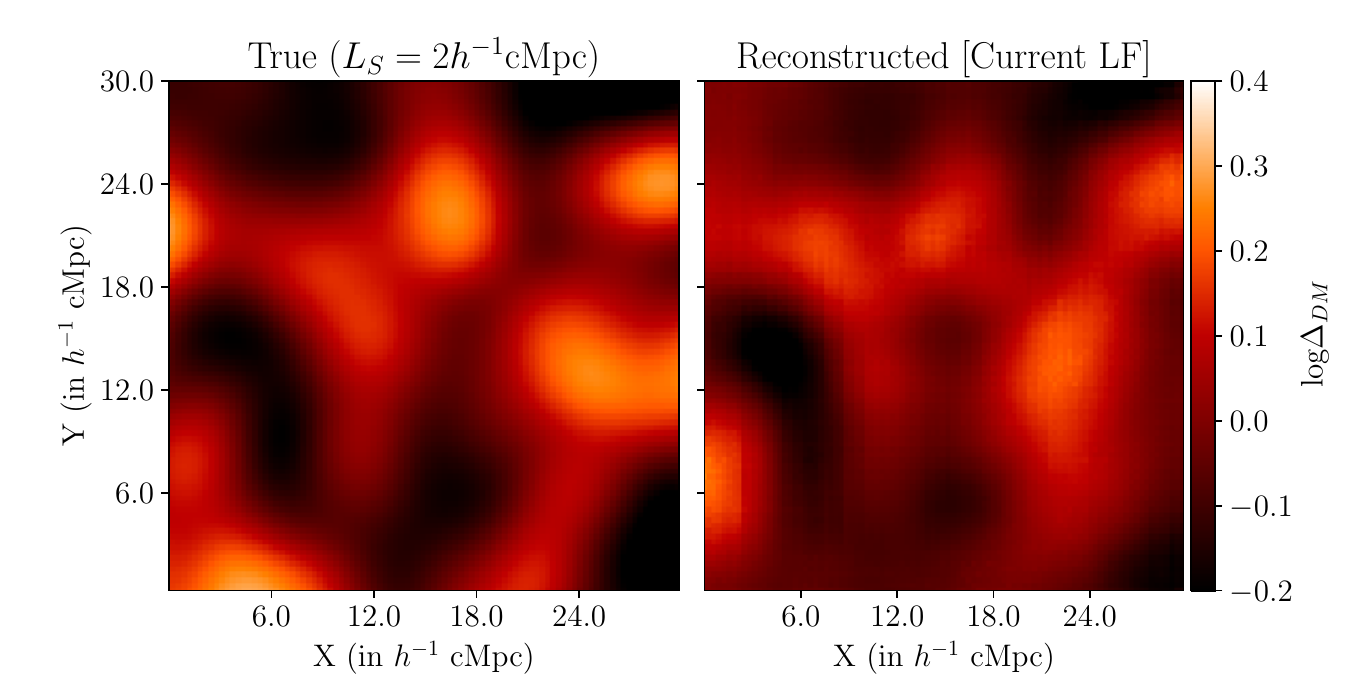}%
    \includegraphics[width=5.3cm, trim={11.6cm 0cm 0cm 0cm}, clip]{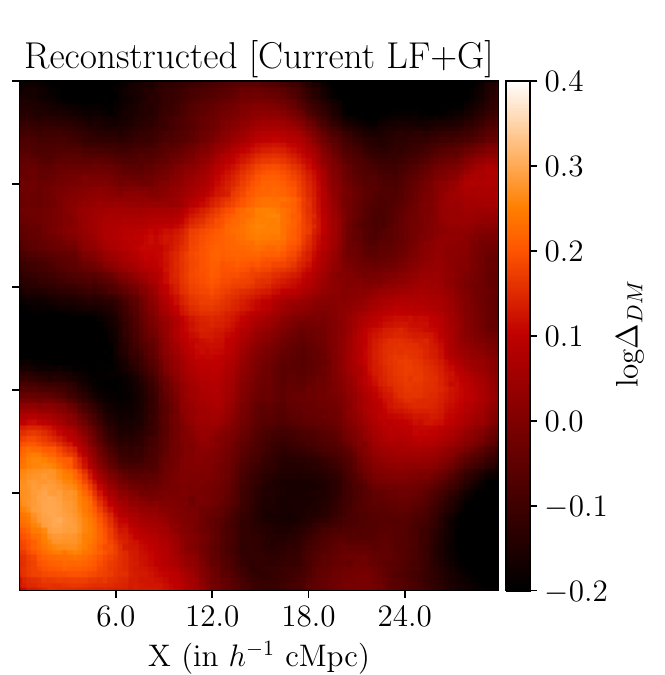}
    \caption{
Slices of thickness $2\,h^{-1}\,\mathrm{cMpc}$ through the 3D dark matter density field. 
\textbf{From left to right:} 
(1) The true unsmoothed $\log \Delta_{\rm DM}$ field with the resolution of the simulation. 
(2) The true $\log \Delta_{\rm DM}$ field smoothed with a Gaussian filter of scale $L_S = 2\,h^{-1}\,\mathrm{cMpc}$. 
(3) Reconstruction of the smoothed density field using only Ly$\alpha$ forest (LF) data. 
(4) Reconstruction of the smoothed density field using combined Ly$\alpha$ forest and galaxy (LF+G) data. 
All reconstructions correspond to the ``current'' survey scenario with sightline separation $2.4\,h^{-1}\,\mathrm{cMpc}$ and a galaxy population number density consistent with such surveys. 
The inclusion of galaxies in training and testing improves the recovery of dense regions and enhances the fidelity of the large-scale structure reconstruction compared to using LF data alone.
}

    \label{fig:tomography}
\end{figure*}

\subsection{Modeling Galaxy Fields} \label{sec:galaxies}

To model realistic galaxy distributions for tomographic reconstruction, we begin by identifying dark matter halos within the simulation using a friends-of-friends (FoF) algorithm with a linking length of 0.2 and a minimum particle threshold of 5. Galaxies are then populated into these halos using a halo occupation distribution (HOD) framework, which statistically maps dark matter halos to galaxy populations. Specifically, we adopt the \citet{zheng2007} model as implemented in the \textsc{halotools} package \citep{hearin2017}, with HOD parameters $\alpha = 0.5$, $\sigma_{\log M} = 0.40$, and a luminosity threshold of $M_r < -18$, assigning both central and satellite galaxies to each halo according to its mass.
From the full HOD catalog, we construct observationally motivated subsamples by uniformly downsampling the galaxy population, treating centrals and satellites with equal weight, to match the expected number densities of different survey regimes. For current surveys such as Subaru-PFS, CLAMATO, and LATIS, we match the number density inferred from observed cumulative luminosity functions, as shown in Figure~\ref{fig:survey}. For future surveys, including ELT/MOSAIC, we adopt a deeper selection corresponding to a $\sim$15-fold increase in number density, consistent with projections in Figure~\ref{fig:survey}.

To account for observational effects, we incorporate redshift-space distortions by displacing galaxies along the line of sight according to their peculiar velocities. For this work, we do not include uncertainties in galaxy redshifts, although such uncertainties, arising from spectroscopic measurement errors, may degrade the fidelity of tomographic reconstruction in real observations.
For training, the resulting galaxy distributions are projected onto three-dimensional grids using cloud-in-cell (CIC) interpolation and are used in conjunction with Ly$\alpha$ forest data as inputs for tomographic reconstruction. The CIC is performed on a \(128 \times 128 \times 384\) grid to maintain consistent input dimensions with the Ly$\alpha$ forest field.

\section{Neural Network Architecture and Training Procedure}\label{sec:3dunet_vae}

We introduce \textsc{DeepCHART}\footnote{\textsc{DeepCHART}: Deep learning for Cosmological Heterogeneity and Astrophysical Reconstruction via Tomography. Code available at \url{https://github.com/soumak-maitra/DeepCHART}.}, a neural network framework for inferring three-dimensional density fields from sparse astrophysical observables. At its core, \textsc{DeepCHART} employs a three-dimensional Variational Autoencoder (VAE) built upon a 3D U-Net architecture \citep{Ronneberger2015, Cicek2016}, designed to capture the probabilistic structure of the inverse mapping from observational tracers to the underlying matter distribution.

To enable efficient training and data augmentation, we extract subvolumes of size \((30\,h^{-1}\mathrm{cMpc})^3\) from a larger simulation box of \((40\,h^{-1}\mathrm{cMpc})^3\). This subvolume sampling exposes the network to a wider variety of cosmic structures and improves generalization by reducing overfitting. All training and inference are performed on \(30\,h^{-1}\mathrm{cMpc}\) subvolumes.

Each \(30\,h^{-1}\mathrm{cMpc}\) subvolume is discretized into a \(96\times96\times288\) grid, chosen to maintain an effective spatial resolution compatible with observations. In particular, the grid is asymmetric along the line-of-sight (\(z\)) direction, with the third dimension having three times as many cells as the transverse directions. This configuration ensures that each voxel along the \(z\)-axis matches the spectral pixel size of real Ly$\alpha$ forest observations, providing a physically meaningful input structure. The Ly$\alpha$ forest input is represented as a sparse 3D box, where absorption sightlines are inserted at regular intervals in the \(x\)–\(y\) plane and extend along the \(z\)-axis. Voxels not intersected by a sightline are set to zero.

To ensure consistent input dimensions, we apply the same gridding scheme to the galaxy field. Galaxy positions are interpolated onto the \(96\times96\times288\) grid using a cloud-in-cell (CIC) algorithm, resulting in a continuous 3D overdensity field aligned with the Ly$\alpha$ input. These two fields, Ly$\alpha$ and galaxy, are stacked to form a two-channel input volume for the neural network.

The encoder employs fully convolutional layers and residual blocks, structured into four convolutional stages with progressively increasing feature channels (16, 32, 64, and 128 channels, respectively). In each stage, convolutional layers use kernels of size \(7\times7\times21\), \(5\times5\times15\), and two \(3\times3\times9\), with the larger extent along the \(z\)-axis to account for the grid anisotropy. Each stage includes ReLU activations, residual connections, and spatial downsampling with a stride of 2, reducing the input volume from \(96\times96\times288\) to a compact latent representation of size \(12\times12\times36\).

This encoded tensor is then flattened and passed through two fully connected layers to compute the latent distribution parameters, mean \(\mu\) and log-variance \(\log\sigma^2\), which define a Gaussian posterior in a 512-dimensional latent space. The log-variance is preferred over directly learning the standard deviation \(\sigma\) to ensure numerical stability and maintain positivity during training. During training, a latent vector \(Z\) is obtained by sampling \(\varepsilon \sim \mathcal{N}(0, I)\) and applying the reparameterization trick, $Z = \mu + \exp\left(\tfrac{1}{2}\log\sigma^2\right)\odot\varepsilon$,
where the exponential maps the log-variance to the standard deviation. This formulation allows unbiased stochastic sampling with end-to-end differentiability \citep{KingmaWelling2013}.

The decoder mirrors the encoder's architecture and performs progressive upsampling via transposed convolutions and residual blocks. It first projects the latent vector \(Z\) into a tensor of shape \(12 \times 12 \times 36 \times 128\). This tensor is then upsampled through four stages to recover the output spatial size of \(96\times96\times96\). Skip connections from the encoder at corresponding scales (e.g., \(12^3\), \(24^3\), \(48^3\), etc.) enrich the decoding path with high-resolution features. Finally, the output convolution maps the last feature map into a scalar field, yielding the reconstructed 3D density contrast field \(\hat{\mathbf{X}} \in \mathbb{R}^{96\times96\times96}\). The entire schematic of the architecture is shown in Figure~\ref{fig:unetvae}.

We consider two survey configurations in this work. In the ``current'' scenario, with a mean sightline separation of \(2.4\,h^{-1}\mathrm{cMpc}\), we reconstruct the dark matter field smoothed on \(2\,h^{-1}\mathrm{cMpc}\) scales. In the ``future'' scenario with \(1.0\,h^{-1}\mathrm{cMpc}\) separation, the model is trained to reconstruct dark matter fields smoothed at both \(2\,h^{-1}\mathrm{cMpc}\) and \(1\,h^{-1}\mathrm{cMpc}\) scales. The galaxy number density is adjusted in each case to reflect the corresponding observational depth and completeness.

Training is performed using the evidence lower bound (ELBO) loss:
\begin{equation}
    L(\theta, \phi) = \mathrm{MSE}(\hat{\mathbf{X}}, \mathbf{X}) + \beta \cdot \mathrm{KL}\big(q_\phi(z|\mathbf{Y}, \mathbf{X}) \| p(z)\big),
\end{equation}
where the reconstruction loss is a mean squared error (MSE) between the predicted and true density fields, and the KL divergence term regularizes the latent distribution. The hyperparameter \(\beta\) follows a \(\beta\)-VAE annealing strategy \citep{Higgins2017}, increasing from 0.1 to 1 up to a training epoch of 100. The prior \(p(z)\) is a standard multivariate Gaussian, \(\mathcal{N}(0, I)\).

The VAE is trained on an NVIDIA V100 GPU using the Adam optimizer \citep{KingmaWelling2013}, with a learning rate of \(10^{-4}\), batch size of 16, and for 500 epochs. Each epoch takes approximately 4 minutes, totaling around 2000 GPU minutes or \(\sim\)33 GPU hours on a single V100 device. The validation loss typically stabilizes around 400 epochs.

At inference time, the encoder maps input observations to the latent parameters (\(\mu\), \(\log \sigma^2\)), from which a latent sample is drawn and decoded to produce a full 3D density reconstruction. Each inference run takes less than 0.5 seconds per subvolume, offering a significant computational speed-up over traditional Bayesian approaches such as TARDIS \citep{Horowitz2019}. The complete architecture is illustrated in Fig.~\ref{fig:unetvae}.

\section{Results}\label{sec:results}
We present the performance of our deep learning tomographic reconstruction on the test dataset. We detail the results for field reconstruction quality (Section~\ref{sec:field_accuracy}), density distribution (Section~\ref{sec:delta_pdf}), power spectrum and two-point statistics (Section~\ref{sec:powerspec}), and the recovery of cosmic web structures (Section~\ref{sec:web}). Example visualizations are also provided to qualitatively illustrate the reconstruction results.

\begin{figure*}
    \centering
  
    \includegraphics[width=13cm, trim={0cm 0cm 0 1cm}, clip]{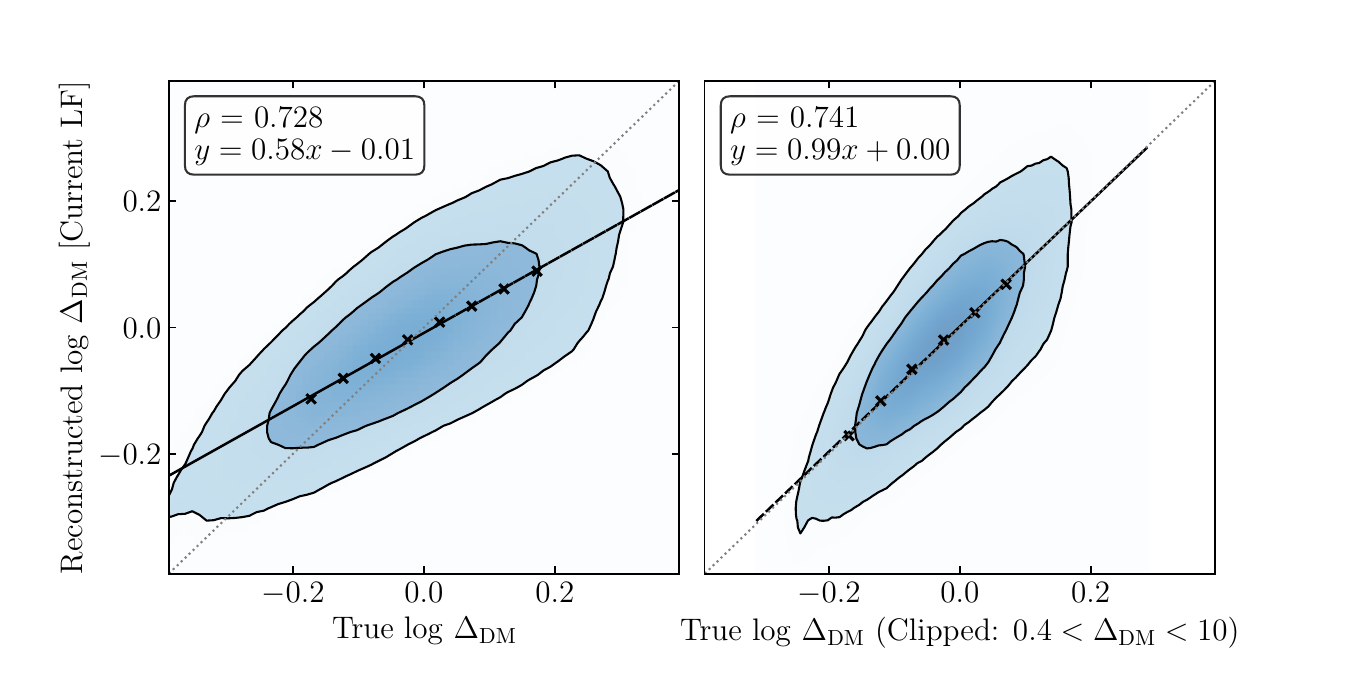}

    \includegraphics[width=13cm, trim={0cm 0cm 0 1cm}, clip]{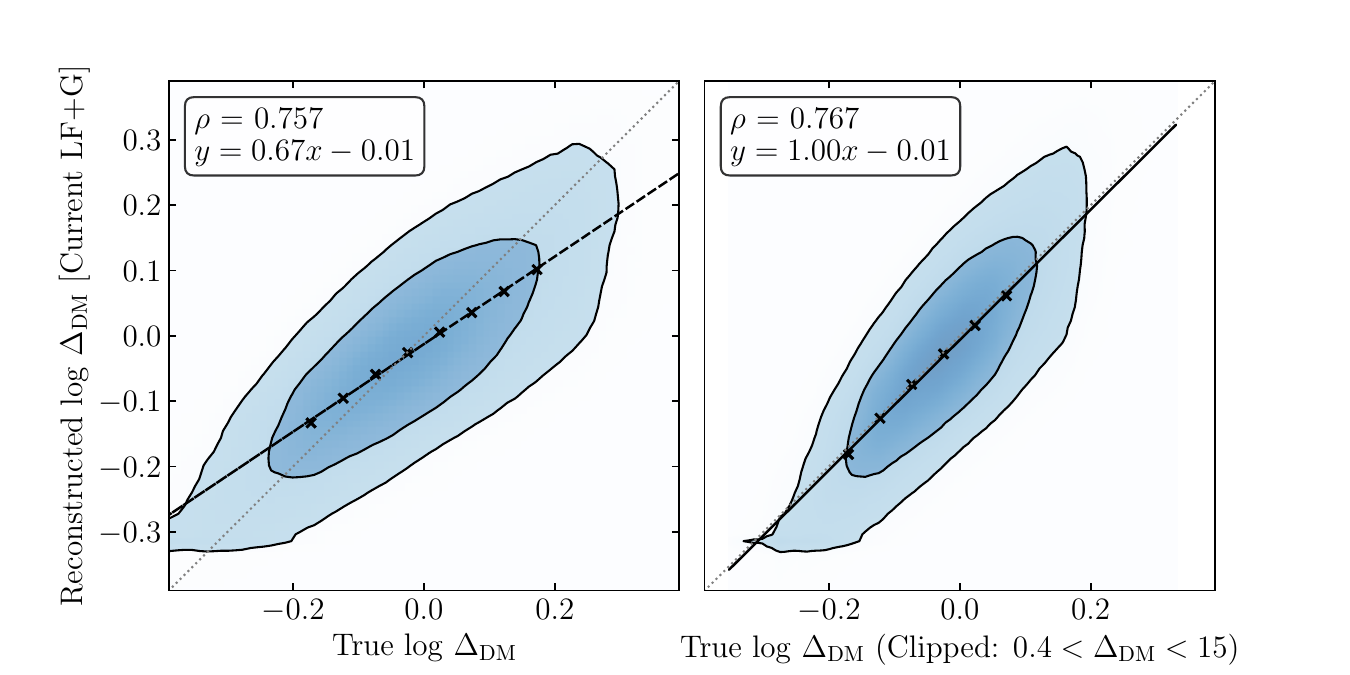}
    \caption{
Voxel-wise comparison between the reconstructed and true $\text{log} \Delta_{\rm DM}$ values from the VAE reconstructions, with the dark matter field smoothed over a scale of $2\,h^{-1}\,\mathrm{cMpc}$. Each panel shows 2D density contours and binned medians of the predictions, along with a robust linear fit and the Pearson correlation coefficient $\rho$. Note that the clipping thresholds are applied based on the original (unsmoothed with simulation resolution) dark matter field, while the plotted values correspond to the smoothed field.
\textbf{Top row:} Model trained and tested using only Ly$\alpha$ forest data. The left panel shows that the dynamic range is limited, with underprediction at both low and high densities. After clipping to $0.4 < \Delta_{\rm DM} < 10$ (right), the correlation improves and the slope approaches unity, indicating that the reconstruction is reliable up to moderate overdensities.
\textbf{Middle row:} Model trained on both Ly$\alpha$ forest and galaxy data, but tested using only Ly$\alpha$ forest. Including galaxies during training improves the reconstruction even when galaxy information is absent at test time. Clipping to $0.4 < \Delta_{\rm DM} < 12$ (right) further improves the correlation and slope.
\textbf{Bottom row:} Model trained and tested using both Ly$\alpha$ forest and galaxy data. The clipped panel (right) shows accurate reconstruction over a wider dynamic range ($0.4 < \Delta_{\rm DM} < 15$), with correlation and slope remaining close to unity, indicating that the inclusion of galaxies extends the fidelity of the reconstruction into higher-density regions.
}
\label{fig:contour_accuracy}
\end{figure*}

\subsection{Reconstruction Accuracy}\label{sec:field_accuracy}

To assess the accuracy of our deep-learning-based tomographic reconstructions, we compare the reconstructed and true dark matter density fields across multiple survey scenarios and data combinations. This includes visual inspection of transverse slices to evaluate the recovery of large-scale structures, and voxel-wise statistical comparisons using Pearson correlation coefficients and linear regression fits to quantify reconstruction fidelity.

\subsubsection{Reconstruction using Current Surveys}

Figure~\ref{fig:tomography} demonstrates the performance of our reconstruction framework, {\sc DeepCHART}, in recovering the dark matter density field at redshift $z=2.5$ from sparse observational tracers. We focus here on conditions representative of current Ly$\alpha$ forest surveys such as Subaru PFS, CLAMATO, and LATIS, which typically achieve a mean sightline separation of $2.4\,h^{-1}\,\mathrm{cMpc}$. This sampling scale naturally limits the smallest structures that can be reliably reconstructed. To enable meaningful comparisons, we smooth the true simulated density field on a matching scale of $L_S = 2\,h^{-1}\,\mathrm{cMpc}$, corresponding to the effective spatial resolution set by the data.

Each panel shows a transverse slice of thickness $2\,h^{-1}\,\mathrm{cMpc}$ through the simulation box. From left to right, we display: (i) the true log-density field with simulation resolution without smoothing, (ii) the smoothed reference field, and (iii) two reconstructions obtained using either only Ly$\alpha$ forest (LF) data or a combination of LF and galaxy (LF+G) tracers. The LF-only reconstruction captures the overall morphology of the large-scale structure, including prominent voids and extended filaments, but tends to underestimate sharp density contrasts and compact features due to the saturation of Ly$\alpha$ absorption in high-density regions, which limits the sensitivity of the forest to those environments.
When galaxy data are incorporated, the reconstruction exhibits improved fidelity, recovering finer filaments and dense nodes. This demonstrates the added value of combining galaxies with forest data, particularly in regions where the Ly$\alpha$ forest becomes less sensitive.

    To quantify the accuracy of the reconstructed fields, we perform a voxel-wise comparison between the predicted and true dark matter log-densities using the Pearson correlation coefficient, defined as
\begin{equation}
    \rho = \frac{\mathrm{Cov}(x, y)}{\sigma_x \sigma_y},
\end{equation}
where $\mathrm{Cov}(x, y)$ is the covariance between the true ($x$) and reconstructed ($y$) log-density fields, and $\sigma_x$, $\sigma_y$ are their respective standard deviations. A value of $\rho = 1$ corresponds to perfect linear correlation, while $\rho = 0$ indicates no correlation.

Quantitative results for the LF-only reconstruction are shown in Figure~\ref{fig:contour_accuracy}. When using the full dynamic range of the density field, we find a Pearson correlation coefficient of $\rho = 0.728$ and a best-fit linear regression slope of $0.58$. The best-fit relation is determined by first identifying the $1\sigma$ confidence contour around the 2D voxel distribution and binning the true log-density values along the $x$-axis in intervals of width $0.05$. Within each bin, we compute the median predicted value, and fit a straight line through these median points to characterize the overall bias in the reconstruction. The resulting sub-unity slope indicates a systematic underestimation of the amplitude of density fluctuations, particularly in the high- and low-density extremes.

This trend is consistent with known limitations of the Ly$\alpha$ forest as a tracer of the matter distribution. In high-density regions, forest absorption saturates, resulting in flattened transmitted flux values that erase sensitivity to fluctuations in the underlying field. In underdense regions, the flux approaches the continuum, where instrumental noise dominates and reduces the reliability of the signal. Additionally, the finite sightline separation in current surveys limits the spatial sampling of the field, making it more difficult to recover small-scale features, particularly in extreme overdensities and voids. Increasing the density of sightlines would alleviate this issue by providing more localized information, thereby improving reconstruction fidelity in both low- and high-density regimes. To mitigate the influence of poorly constrained extremes in the current setup, we apply a clipping procedure to the density field, restricting the comparison to an intermediate overdensity range where the forest signal remains most informative.
After clipping to $0.4 < \Delta_{\mathrm{DM}} < 10$, the correlation coefficient improves to $\rho = 0.741$, and the best-fit slope approaches unity, $y = 0.99x$. This indicates a more faithful reconstruction in the moderately overdense regime, where the Ly$\alpha$ forest remains most sensitive.

The combined LF+G reconstruction (bottom row, Figure~\ref{fig:contour_accuracy}) yields consistently higher correlation, reaching $\rho = 0.757$ without clipping and $\rho = 0.767$ after applying a broader density cut of $0.4 < \Delta_{\mathrm{DM}} < 15$. In both cases, the slope of the fit remains effectively unity ($y = 1.00x$), indicating that the reconstruction accurately preserves the amplitude of fluctuations across a wider dynamic range. The inclusion of galaxy data enhances the recovery of high-density regions where the forest alone becomes insensitive, highlighting the benefit of combining multiple tracers. These results demonstrate the flexibility and effectiveness of {\sc DeepCHART} in jointly assimilating Ly$\alpha$ forest and galaxy observations to reconstruct the underlying matter distribution with improved accuracy. 

We further explore in Appendix~\ref{A:Cross-Modal} how such models, even when applied to forest-only data at inference time, outperform those trained solely on forest inputs. The modest but consistent improvement in correlation coefficient and dynamic range ($0.4 \leq \Delta_{\mathrm{DM}} \leq 12$; see Figure~\ref{fig:lfonly_infer}) highlights the model’s ability to retain knowledge of high-density mappings acquired during training. This underscores the value of joint tracer training even when only partial data are available at test time.

\begin{figure*}
    \centering
    \noindent\textbf{Future Tomography with Dark Matter Density smoothed over $L_S=2h^{-1}$cMpc}
    \fbox{%
        \parbox{\textwidth}{%
            \centering
            \includegraphics[width=0.52\textwidth, trim={1cm 0 0 0.5cm}, clip]{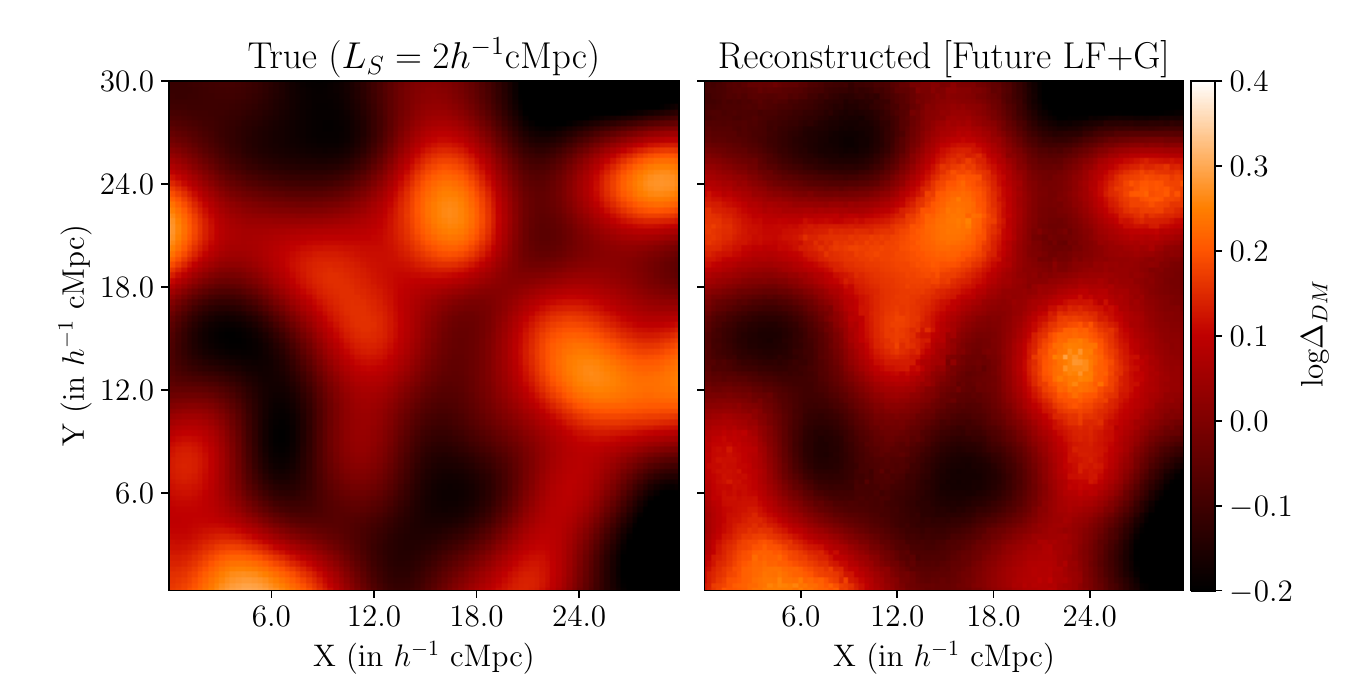}%
            \includegraphics[width=0.48\textwidth, trim={1cm 0 1.8cm 0.5cm}, clip]{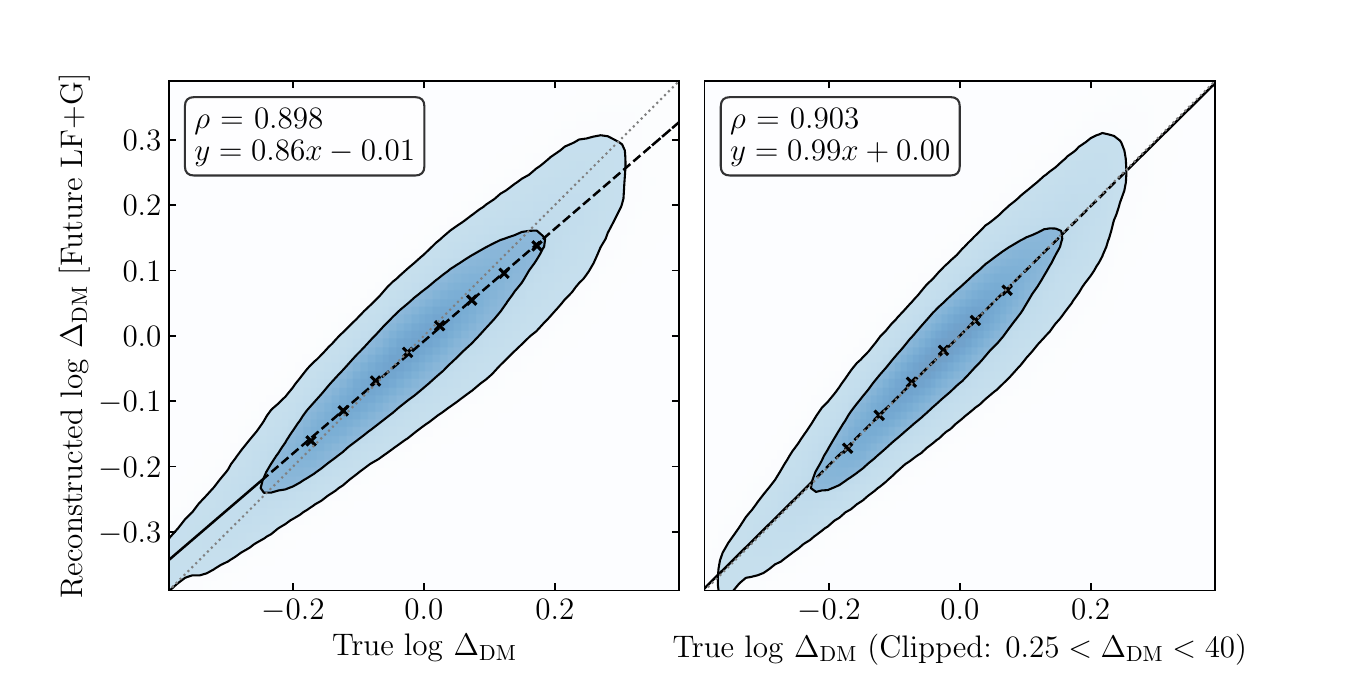}%
        }%
    }

    \vspace{1em}

    \noindent\textbf{Future Tomography with Dark Matter Density smoothed over $L_S=1h^{-1}$cMpc}
    \fbox{%
        \parbox{\textwidth}{%
            \centering
            \includegraphics[width=0.52\textwidth, trim={1cm 0 0 0.5cm}, clip]{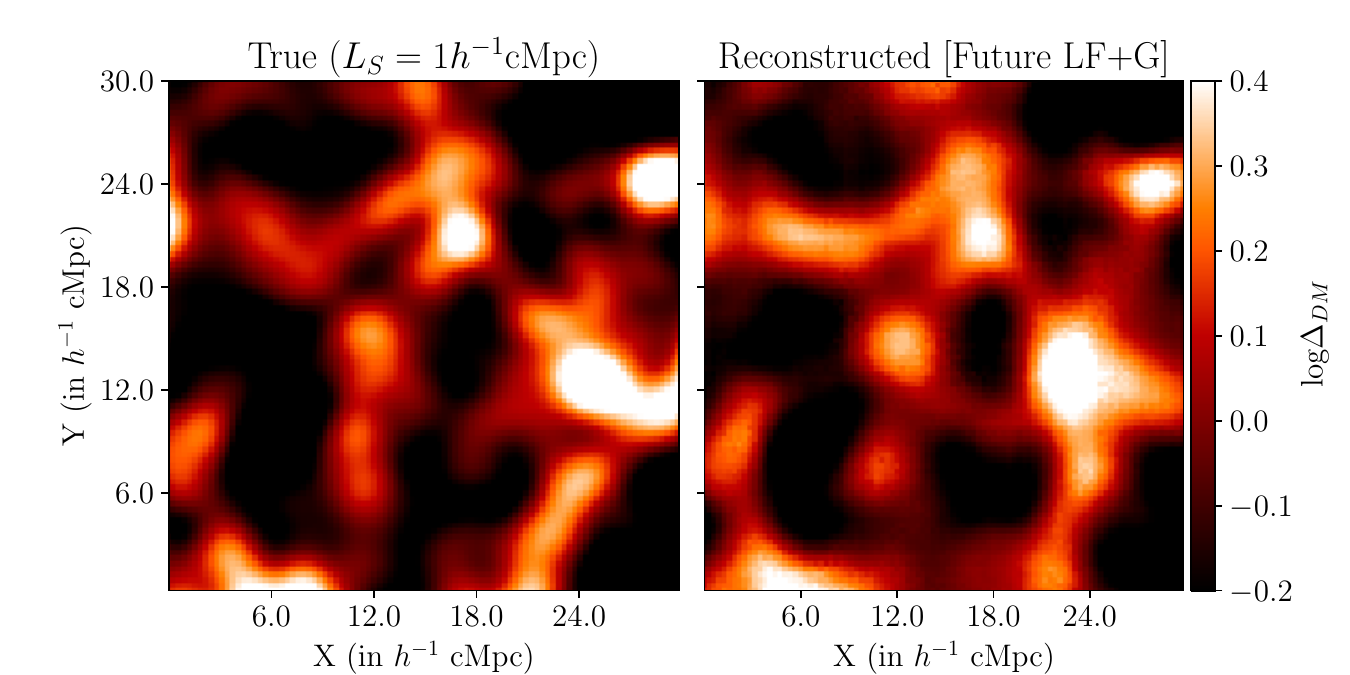}%
            \includegraphics[width=0.48\textwidth, trim={1cm 0 1.8cm 0.5cm}, clip]{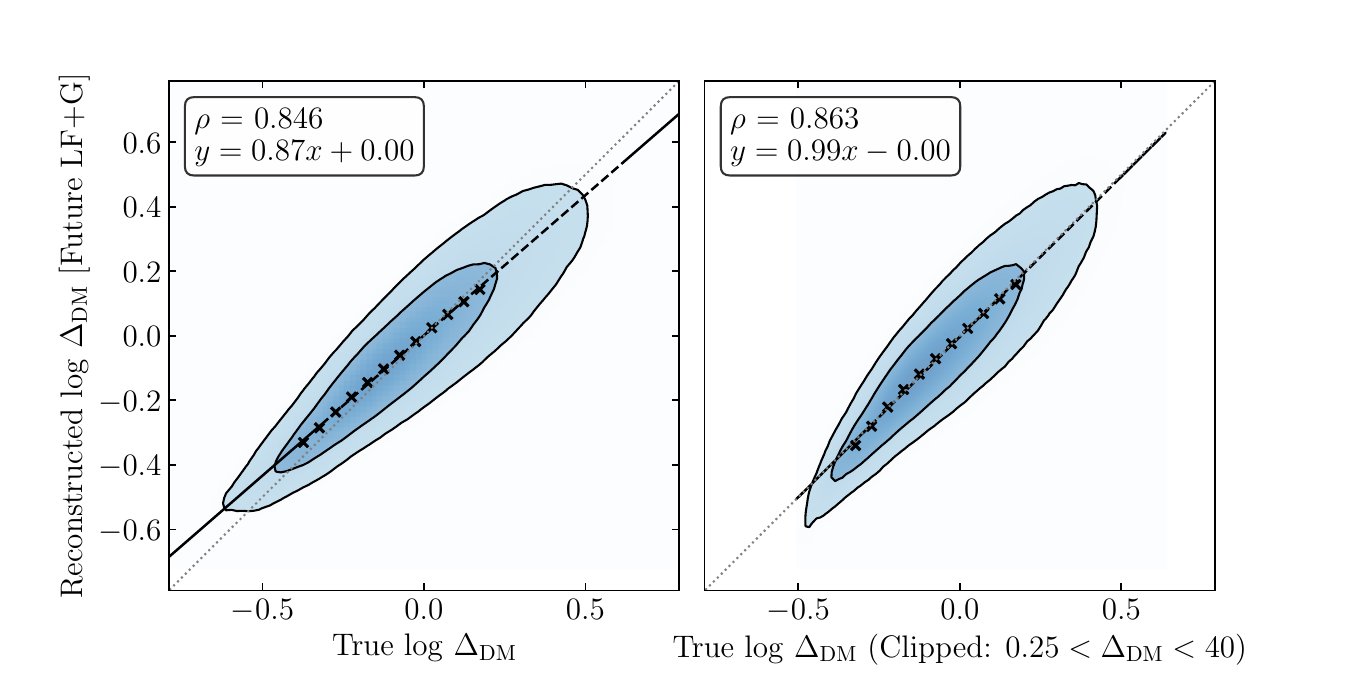}%
        }%
    }
    \caption{
Slices and voxel-wise comparisons of the reconstructed and true dark matter density fields at \(z=2.5\). LF+G denotes reconstructions combining both Ly$\alpha$ forest sightlines and galaxy tracers.
\textbf{Top row:} 2D slices of the true and reconstructed fields, shown in \(\log \Delta_{\rm DM}\), smoothed over \(2\,h^{-1}\mathrm{cMpc}\) with slice thickness of \(2\,h^{-1}\mathrm{cMpc}\). The reconstruction uses mock observations from a hypothetical future survey combining Ly\(\alpha\) forest sightlines with mean separation of \(1\,h^{-1}\mathrm{cMpc}\) and a galaxy sample five times denser than Subaru PFS expectations. The first and second panels show spatial slices of the true and reconstructed fields, respectively. The third and fourth panels show voxel-wise scatter plots comparing reconstructed and true \(\log \Delta_{\rm DM}\) values; the third panel includes all voxels, while the fourth panel is restricted to moderate overdensities (\(0.25 < \Delta_{\rm DM} < 40\)).
\textbf{Bottom row:} Same as top, but with smoothing and slice thickness of \(1\,h^{-1}\mathrm{cMpc}\). The first and second panels show the true and reconstructed spatial slices, and the third and fourth panels show the voxel-wise scatter plots for all voxels and clipped overdensities, respectively.
In both rows, blue contours indicate 68\% and 95\% confidence regions in the scatter plots, and the black lines show robust linear fits, with Pearson correlation coefficients \(\rho\) and regression equations annotated. The reconstruction quality is high, with slopes near unity and strong correlations, particularly in the clipped overdensity range.
}

    \label{fig:tomography_future}
\end{figure*}

\subsubsection{Reconstruction using Future Surveys}

We now turn to future survey scenarios that promise significantly improved sampling of the intergalactic medium. As shown in Figure~\ref{fig:survey}, one can expect upcoming instruments such as WST and ELT/MOSAIC to achieve mean sightline separations of approximately $1.3\,h^{-1}\,\mathrm{cMpc}$ and $1.0\,h^{-1}\,\mathrm{cMpc}$, respectively. Motivated by this, we adopt a mean separation of $1\,h^{-1}\,\mathrm{cMpc}$ pertaining to future survey scenarios to evaluate the potential gains in tomographic reconstruction with higher-resolution data. In this context, we consider reconstructions that jointly utilize Ly$\alpha$ forest sightlines and galaxy positions, reflecting the multi-tracer capabilities anticipated from next-generation spectroscopic surveys.
We explore two complementary questions in this regime. Firstly, we assess how our previous results, based on reconstructions smoothed at $L_S = 2\,h^{-1}\,\mathrm{cMpc}$ to match current survey limits, are improved when applied to a denser sightline configuration. Secondly, we investigate the reconstruction fidelity at a finer smoothing scale of $L_S = 1\,h^{-1}\,\mathrm{cMpc}$, which becomes accessible with such high-density sampling. At these scales, complex baryonic effects and non-linearities become increasingly important, and our use of {\sc DeepCHART}, trained on hydrodynamic simulations, allows us to incorporate this complexity into the reconstruction framework. These comparisons enable us to quantify not only the gains in recovering large-scale structures but also the extent to which small-scale features in the cosmic web can be reliably reconstructed with next-generation surveys. The corresponding visual and statistical results are presented in Figure~\ref{fig:tomography_future}.

For the $L_S = 2\,h^{-1}\,\mathrm{cMpc}$ scale (top row, Figure~\ref{fig:tomography_future}), the reconstructed field closely follows the true smoothed density field, capturing both diffuse large-scale structures and sharper high-density peaks with significantly improved accuracy. Quantitatively, the voxel-wise correlation reaches $\rho = 0.898$, with a best-fit linear relation of $y = 0.86x$ when no clipping is applied. After clipping to the overdensity range $0.25 < \Delta_{\mathrm{DM}} < 40$, the correlation improves slightly to $\rho = 0.903$, and the slope converges to near-unity, $y = 0.99x$.
These results demonstrate that reconstructions under future survey conditions recover not only the spatial morphology but also the amplitude of fluctuations more reliably than in current survey scenarios. The broader clipping range used here reflects the ability to recover a wider dynamic range in the density field reconstruction than what was feasible before, due to the increased sightline sampling and tracer density. While some limitations in extremely overdense or underdense regions may still persist, the overall improvement highlights the potential of high-density spectroscopic surveys to enable more complete tomographic mapping of the cosmic web.

Reducing the smoothing scale to $L_S = 1\,h^{-1}\,\mathrm{cMpc}$ (bottom row, Figure~\ref{fig:tomography_future}) naturally challenges reconstruction accuracy due to the higher resolution demand. Nevertheless, the reconstructed fields continue to exhibit excellent structural fidelity, successfully identifying finer filamentary networks and more localized density enhancements. Quantitatively, we find a correlation coefficient of $\rho = 0.846$ without clipping, improving further to $\rho = 0.863$ after clipping. The slope of the regression remains very close to unity ($y = 0.99x$), underscoring the method's robustness at small scales enabled by high-density sightline and galaxy data. Thus, with the advent of future observational surveys, the ability to reconstruct smaller-scale cosmic web structures with high accuracy will substantially advance our understanding of cosmic structure formation at high redshift.

\begin{figure*}
    \centering
    \includegraphics[width=16.5cm, trim={0cm 0 0cm 0cm}, clip]{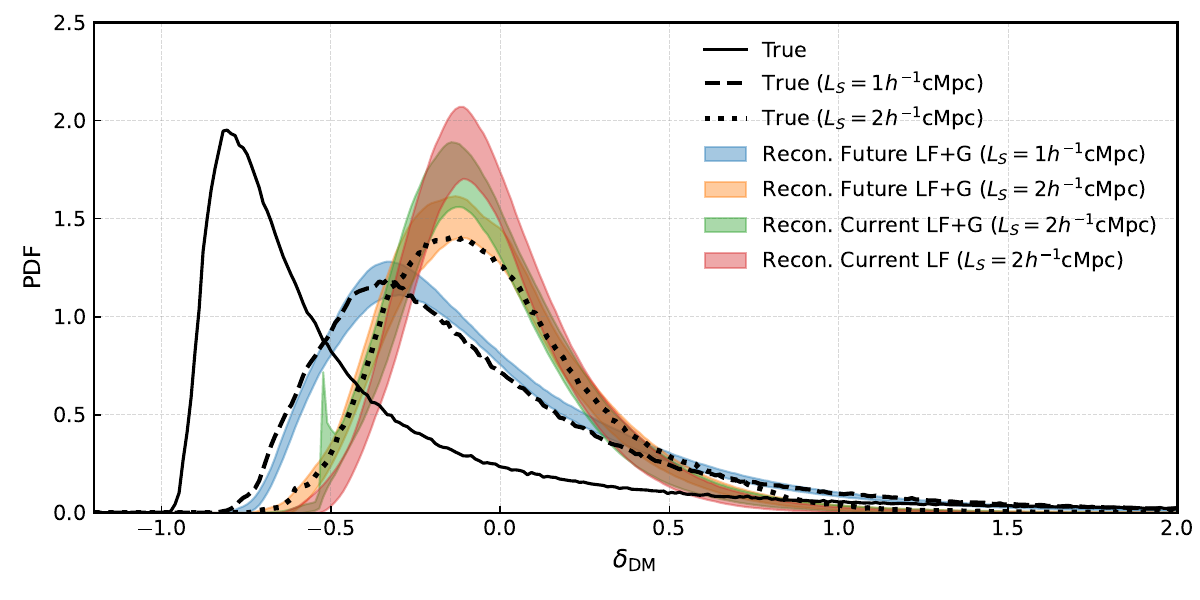}
    \caption{One-point probability distribution functions (PDFs) of the dark matter density contrast, $\delta_{\mathrm{DM}} = \rho / \bar{\rho} - 1$, comparing the true simulated distributions (solid, dashed, and dotted black lines) to reconstructed fields from {\sc DeepCHART} under different survey configurations and smoothing scales. Results are shown for the true unsmoothed field with simulation resolution and for fields smoothed at $L_S = 1~h^{-1}$cMpc and $2~h^{-1}$cMpc. Reconstructed PDFs are plotted for both current and future survey regimes, using Ly$\alpha$ forest-only (LF) for current and combined Ly$\alpha$ forest and galaxy (LF+G) data for both current and future surveys. Shaded regions denote the standard deviation across 80 test realizations. All reconstructions recover the bulk of the PDF well, with the LF+G models providing notably better agreement in the high-density tail. The increased tracer density in the future configuration enables improved fidelity across the full dynamic range of $\delta_{\mathrm{DM}}$.}
    
    \label{fig:pdf_comparison}
\end{figure*}

\begin{figure*}
    \centering
    \includegraphics[width=16.5cm, trim={0.4cm 1cm 1cm 1.2cm}, clip]{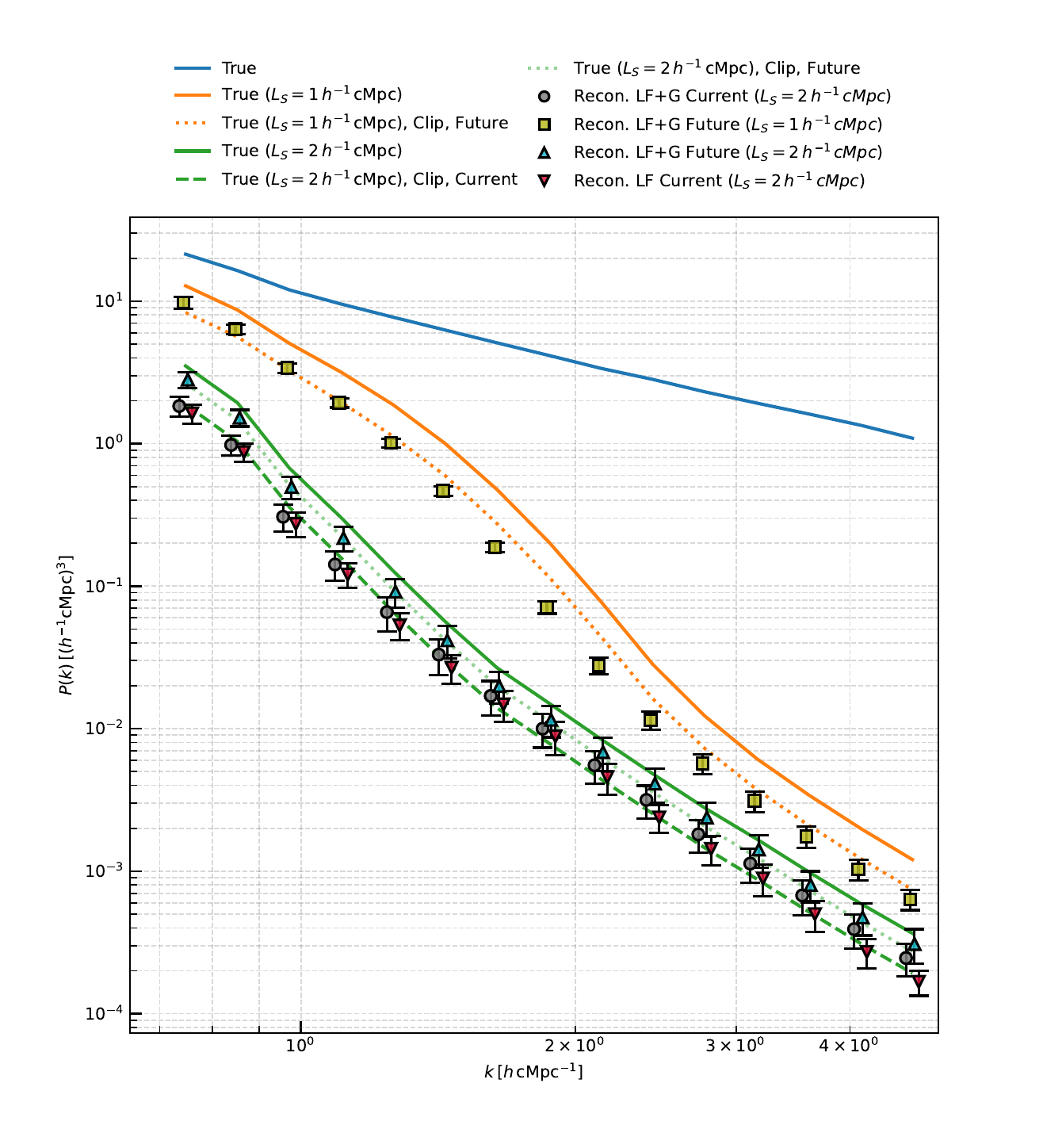}
\caption{
Comparison of the spherically-averaged 3D power spectrum, $P(k)$, for the true dark matter density fields and various reconstructions at redshift $z=2.5$.
The true power spectra from the original simulations, including different smoothing scales ($L_S=1\,h^{-1}\mathrm{cMpc}$ and $L_S=2\,h^{-1}\mathrm{cMpc}$), and clipping applied to mimic survey limitations, are plotted. For future surveys, clipping is performed on overdensities $\Delta$ in the range $0.25 < \Delta < 40$, while for current surveys, clipping uses $0.4 < \Delta < 15$. The reconstructions include multiple models with mean sightline separations corresponding to $1\,h^{-1}\mathrm{cMpc}$ and $2\,h^{-1}\mathrm{cMpc}$. The labels `LF' denote reconstructions using only Ly$\alpha$ forest data, whereas `LF+G' denotes reconstructions with joint Ly$\alpha$ forest and galaxy data. The reconstructed spectra for the current and future surveys are compared to their respective True clipped power spectra. Error bars represent the standard deviation over multiple realizations with varying $(30\,h^{-1}\mathrm{cMpc})^{3}$ subvolumes extracted from our $(40\,h^{-1}\mathrm{cMpc})^{3}$ simulation box, as well as different sightline configurations. Slight horizontal shifts in $k$ are applied to improve visual clarity.
}
    \label{fig:powerspec}
\end{figure*}

\subsection{One-Point Statistics of the Density Field}\label{sec:delta_pdf}

To complement the voxel-wise correlation analysis in Section~\ref{sec:field_accuracy}, we now examine the one-point probability distribution function (PDF) of the dark matter density contrast, $\delta_{\mathrm{DM}} = \rho / \bar{\rho} - 1$. The one-point PDF captures the full non-Gaussian structure of the matter field in a compact form and provides a sensitive probe of reconstruction fidelity across the full dynamic range of densities, including the rare low- and high-density regions. The PDF measures the frequency distribution of voxel values and thus reflects the ability of the network to reproduce the correct amplitude distribution of the underlying field.

Figure~\ref{fig:pdf_comparison} shows the one-point PDFs of the true dark matter field and the reconstructed fields across different survey regimes and tracer configurations. We present results for the current survey configuration (mean sightline separation $d_\perp = 2.4h^{-1}$cMpc) smoothed at $L_S = 2h^{-1}$cMpc, and for the future scenario ($d_\perp = 1.0h^{-1}$cMpc) smoothed at both $L_S = 1h^{-1}$cMpc and $2~h^{-1}$cMpc. The true PDF is shown for both the unsmoothed field with simulation resolution and the smoothed field, while shaded bands represent the standard deviation across 80 realizations of reconstructed fields. The reconstructions include models trained and tested using only Ly$\alpha$ forest (LF; for current survey only), as well as those incorporating both LF and galaxy (LF+G) data.

The true $\delta_{\mathrm{DM}}$ field exhibits a characteristic non-Gaussian shape with a long positive tail, reflecting the skewed nature of matter distribution in the nonlinear regime. The PDF peaks at underdense regions ($\delta \sim -0.8$) and extends out to $\delta \gtrsim 2$ for overdense voxels. Smoothing the field at 1 or 2~$h^{-1}$cMpc reduces the amplitude of fluctuations and contracts the dynamic range, leading to a narrower and more symmetric distribution. This provides a stringent test for the reconstructions, especially in recovering the correct frequency of extreme events.

For the current survey configuration at $L_S = 2h^{-1}$cMpc, the reconstructed PDFs exhibit a significantly narrower distribution compared to the true PDF smoothed at $L_S = 2h^{-1}$cMpc, with most of the probability density concentrated near the mean density ($\delta \sim 0$). This behavior reflects the model's limited ability to accurately reconstruct both low-density voids and high-density structures. The LF-only model shows pronounced overprediction of the probability distribution near $\delta \sim 0$. This trend is consistent with the known limitations of the Ly$\alpha$ forest, which becomes saturated in overdense regions and noisy in underdense ones. The addition of galaxy tracers (LF+G) leads to a modest improvement, with a slightly broader distribution, but still does not fully capture the extremes well. These systematic biases mirror the trends observed in the voxel-wise correlation analysis (Figure~\ref{fig:contour_accuracy}), indicating that the reconstruction remains most reliable in regions of moderate density.

In the future survey scenario with dense tracer sampling ($d_\perp = 1.0h^{-1}$cMpc) and the inclusion of both Ly$\alpha$ forest and galaxy information (LF+G), the reconstructed PDF at $L_S = 2h^{-1}$cMpc shows markedly improved fidelity to the true underlying distribution. The reconstruction closely follows the true PDF, with reduced scatter across realizations. This improvement stems from both the increased number of sightlines and the additional sensitivity of galaxy tracers to overdense regions, which helps partially recover the high-density tail. However, some suppression of the extremely high-density regions remains, consistent with the general bias toward mean-density values observed in all reconstructions.

At finer smoothing ($L_S = 1h^{-1}$cMpc), the true PDF becomes significantly broader and more skewed, reflecting enhanced small-scale structure. The reconstructed distribution, while still capturing the overall shape, exhibits small deviations from the true field. The low-density tail ($\delta \lesssim -0.6$) is slightly suppressed, and the central peak is shifted slightly towards $\delta \sim 0$, again highlighting the model's bias toward mean-density regions when reconstructing at higher resolution. Nonetheless, compared to the current survey results, the future LF+G configuration maintains a better recovery of the high-density tail even at $1h^{-1}$cMpc, suggesting that higher tracer density plays a crucial role in constraining small-scale non-Gaussian features of the dark matter field.

The one-point statistics offer a direct probe of how well the reconstruction preserves the full distribution of matter density contrasts. These results confirm that while the reconstructed fields tend to be biased toward mean densities, the bulk of the PDF can still be recovered with reasonable accuracy, particularly in regimes with higher tracer sampling. Deviations in the low- and high-density tails reflect the known difficulty of recovering extreme environments—deep voids suffer from limited observational sensitivity, while dense peaks are often underrepresented due to saturation and sparse sampling. Nevertheless, improvements seen with increased tracer density suggest that {\sc DeepCHART} is capable of encoding non-Gaussian features of the field when provided with sufficient input information. The ability to match the overall shape and spread of the true PDF underlines that the model does not simply learn smooth morphology, but captures meaningful amplitude variations, thereby retaining key aspects of the nonlinear matter distribution important for understanding cosmic structure formation.

\subsection{Power Spectrum and Large-Scale Clustering}\label{sec:powerspec}

To assess the statistical properties of the reconstructed density fields beyond voxel-wise comparisons or one-point probability distributions, we compute the three-dimensional spherically averaged matter power spectrum, $P(k)$, using the {\sc Pylians} library \citep{Pylians}. Unlike pointwise correlation metrics, the power spectrum characterizes the two-point clustering of matter as a function of scale, capturing how density fluctuations are distributed across different length scales. It thus provides a physically meaningful test of whether the reconstructed fields preserve the underlying statistical structure of the cosmic matter distribution, as expected from large-scale structure formation models.

We perform the Fourier analysis on the dark matter overdensity field, $\Delta_{\mathrm{DM}}(\mathbf{x})$, as reconstructed by {\sc DeepCHART}. While the variational autoencoder is trained on the logarithmic transformation of the density field to improve learning across the wide dynamic range of densities, the power spectrum analysis is carried out on the linear overdensity field to ensure consistency with standard cosmological statistics. This allows a direct comparison with the true matter power spectrum and better reflects the physical clustering of mass in the Universe.
The real-space density fields are discretized onto a cubic grid with dimensions $96^3$, spanning a comoving volume of $(30~h^{-1}\mathrm{cMpc})^3$. We compute the spherically averaged power spectrum $P(k)$ using 15 logarithmically spaced bins in wavenumber space, with $k$ ranging from $0.7$ to $5.0~h\,\mathrm{cMpc}^{-1}$. This $k$-range is chosen to lie within the resolution limits set by the box size, grid spacing, and sightline separation in our tomographic setup. The logarithmic binning ensures balanced sampling of modes across both large scales ($k \sim 0.7~h\,\mathrm{cMpc}^{-1}$) and small scales ($k \sim 5.0~h\,\mathrm{cMpc}^{-1}$).

The $P(k)$ values are computed by averaging the squared modulus of Fourier coefficients, $|\tilde{\delta}(\mathbf{k})|^2$, within spherical shells of constant $k$. We carry out this calculation separately for the reconstructed density field, the original true simulated field, and, importantly, the true simulated density field clipped to a physically motivated overdensity range that reflects the reliable reconstruction regime in each survey scenario (see Figures~\ref{fig:contour_accuracy} and \ref{fig:tomography_future}): $0.4 < \Delta_{\mathrm{DM}} < 15$ for current surveys and $0.25 < \Delta_{\mathrm{DM}} < 40$ for future surveys. This clipping ensures that the comparison focuses on the portions of the field where the reconstructions are expected to be most accurate.

The statistical robustness of our results is enhanced by averaging over multiple realizations. Specifically, we compute power spectra across 20 distinct sub-volumes of size $(30~h^{-1}\mathrm{cMpc})^3$, each extracted from larger $(40~h^{-1}\mathrm{cMpc})^3$ simulation boxes. These sub-volumes partially overlap and sample different spatial regions, allowing us to probe cosmic variance. Additionally, each realization incorporates different observational configurations, including varying arrangements of Ly$\alpha$ forest sightlines and galaxy positions. This ensemble approach ensures that our power spectrum statistics capture both spatial and observational variability, providing a robust estimate of reconstruction performance. 
Figure~\ref{fig:powerspec} presents the resulting power spectra, where reconstructed fields from both current and future survey scenarios are plotted alongside the corresponding clipped and unclipped true density fields, providing a direct visual assessment of reconstruction accuracy. The errors on the reconstructed spectra reflect the standard deviation across these 20 realizations, quantifying the statistical uncertainty in the reconstruction across different sub-volumes and tracer configurations.

Figure~\ref{fig:powerspec} provides a quantitative assessment of the ability of the reconstructed fields to recover the clustering properties of the underlying dark matter distribution, as characterized by the spherically averaged power spectrum, $P(k)$. The plot presents the true power spectrum from the original simulation (solid blue), along with smoothed versions corresponding to $L_S = 2h^{-1}\mathrm{cMpc}$ and $1h^{-1}\mathrm{cMpc}$, reflecting the spatial resolutions achievable with current and future survey configurations, respectively. For each smoothing scale, a clipped true spectrum is also shown to mimic survey observability limits.

Focusing first on the $L_S = 2h^{-1}\mathrm{cMpc}$ case relevant to current surveys, reconstructions using Ly$\alpha$ forest data alone (red downward triangles) slightly underpredict the large-scale power by approximately $20\%$ at $k \sim 1h\mathrm{cMpc}^{-1}$. However, across most of the $k$-range, both the shape and amplitude of the spectrum are recovered within the 1$\sigma$ error bars. Adding galaxy data (grey circles) improves the agreement at large scales, bringing the recovered power closer to the true clipped spectrum. At smaller scales ($k \gtrsim 1.5h\mathrm{cMpc}^{-1}$), a modest overestimation of power by $\sim 20\%$ is observed, which remains within statistical uncertainties.

In the same smoothing regime ($L_S = 2h^{-1}\mathrm{cMpc}$), we observe a notable improvement under future survey conditions. The LF+G reconstruction (cyan upward triangles) closely follows the clipped true spectrum, with the power remaining near the true values across the full range of scales $0.7 \lesssim k \lesssim 5~h\mathrm{cMpc}^{-1}$. This consistency reflects the improved fidelity enabled by increased sightline density and higher tracer sampling expected from future spectroscopic surveys, allowing accurate recovery of the clustering signal even at smaller scales.

We further evaluate reconstruction fidelity at $L_S = 1h^{-1}\mathrm{cMpc}$ using the high-density tracer setup corresponding to future surveys. In this case, the LF+G reconstruction (yellow square markers) successfully recovers the power spectrum at large scales ($k \lesssim 1h\mathrm{cMpc}^{-1}$) and captures the overall spectral shape across the full $k$-range. The power is moderately underpredicted by up to $\sim 40\%$ at intermediate scales around $k \sim 2~h\mathrm{cMpc}^{-1}$, and by $\sim 20\%$ at small scales ($k \gtrsim 3~h\mathrm{cMpc}^{-1}$), though the latter remains at about $1\sigma$ from the mean across realizations. These deviations are not unexpected given the increased nonlinearity and the growing influence of complex baryonic physics at smaller scales. Overall, the reconstruction demonstrates good agreement in both shape and amplitude, indicating that clustering information can be reliably recovered down to $\sim 1h^{-1}\mathrm{cMpc}$ smoothing scales under future survey conditions.

Taken together, these results highlight that {\sc DeepCHART} is capable of reconstructing the clustering statistics of the matter field over a wide dynamic range with high fidelity. Under current survey conditions, large- and intermediate-scale features are already well recovered at $L_S = 2h^{-1}\mathrm{cMpc}$ smoothing. With future survey configurations, reconstructions extend to finer spatial scales, showing promising agreement down to $L_S = 1h^{-1}\mathrm{cMpc}$, where the power spectrum is reproduced with reasonable accuracy across most modes. This performance reflects the model’s ability to learn non-linear structure–tracer relationships from sparse observational data, including the influence of baryonic effects present in the hydrodynamic simulations used for training. Overall, the power spectrum analysis demonstrates that {\sc DeepCHART} can robustly recover two-point clustering statistics across observational regimes, making it a valuable tool for extracting cosmological information from current and forthcoming spectroscopic surveys.

\begin{figure*}
    \centering
    \noindent\textbf{Current Tomography with Dark Matter Density smoothed over $L_S=2h^{-1}$cMpc}
    \fbox{%
        \parbox{\textwidth}{%
            \centering
            \includegraphics[width=0.58\textwidth, trim={1cm 0 0 0.5cm}, clip]{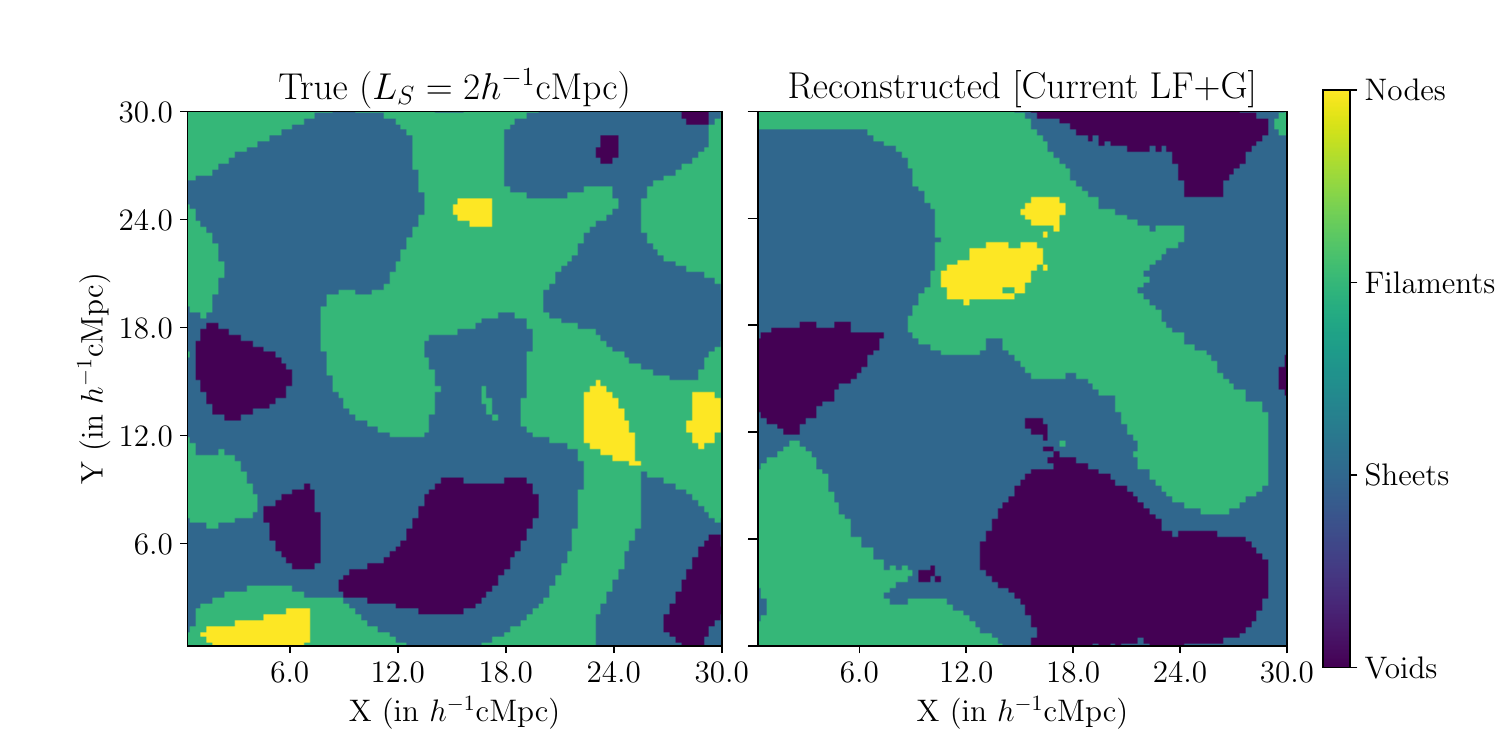}%
            \includegraphics[width=0.4\textwidth, trim={0cm 1.1cm 0cm 0.5cm}, clip]{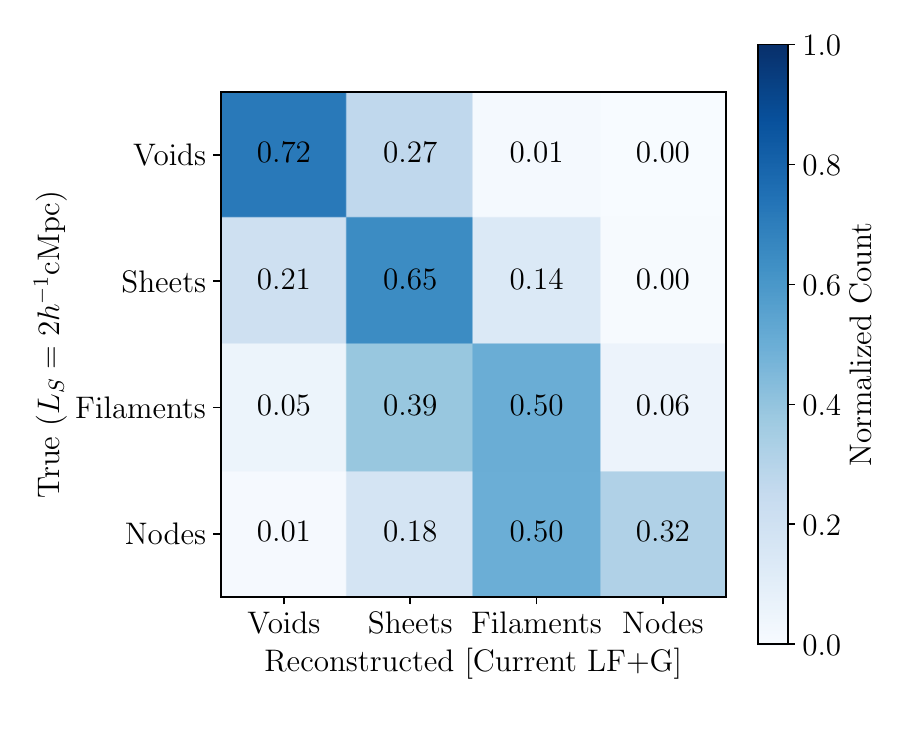}%
        }%
    }

        \vspace{1em}

    \noindent\textbf{Future Tomography with Dark Matter Density smoothed over $L_S=2h^{-1}$cMpc}
    \fbox{%
        \parbox{\textwidth}{%
            \centering
            \includegraphics[width=0.58\textwidth, trim={1cm 0 0 0.5cm}, clip]{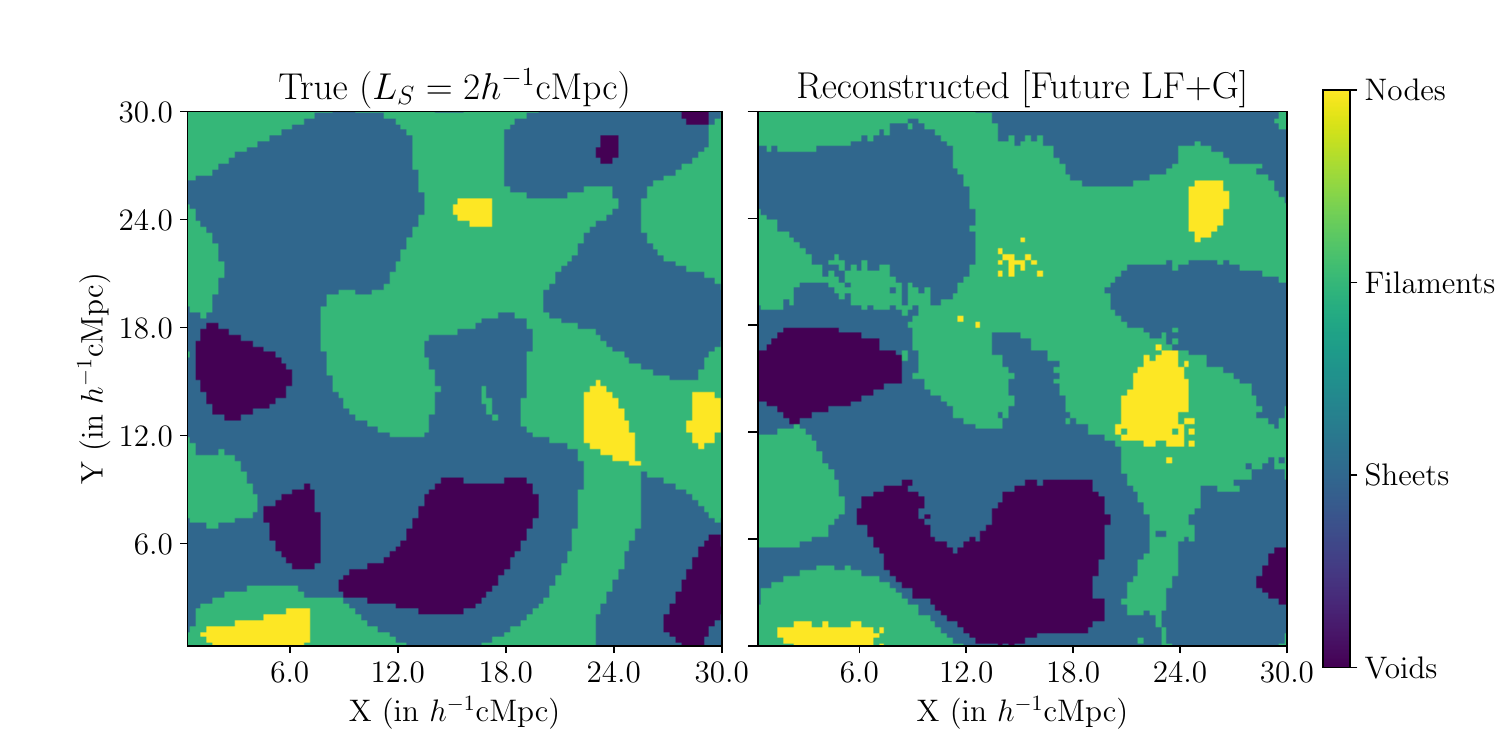}%
            \includegraphics[width=0.4\textwidth, trim={0cm 1.1cm 0cm 0.5cm}, clip]{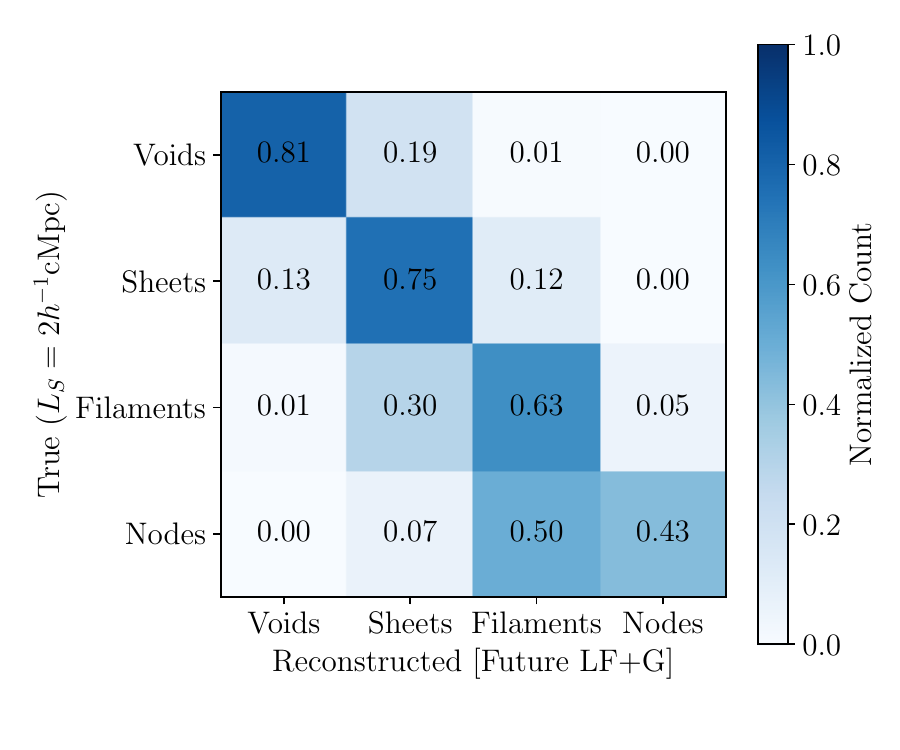}%
        }%
    }

    \vspace{1em}

    \noindent\textbf{Future Tomography with Dark Matter Density smoothed over $L_S=1h^{-1}$cMpc}
    \fbox{%
        \parbox{\textwidth}{%
            \centering
            \includegraphics[width=0.58\textwidth, trim={1cm 0 0 0.5cm}, clip]{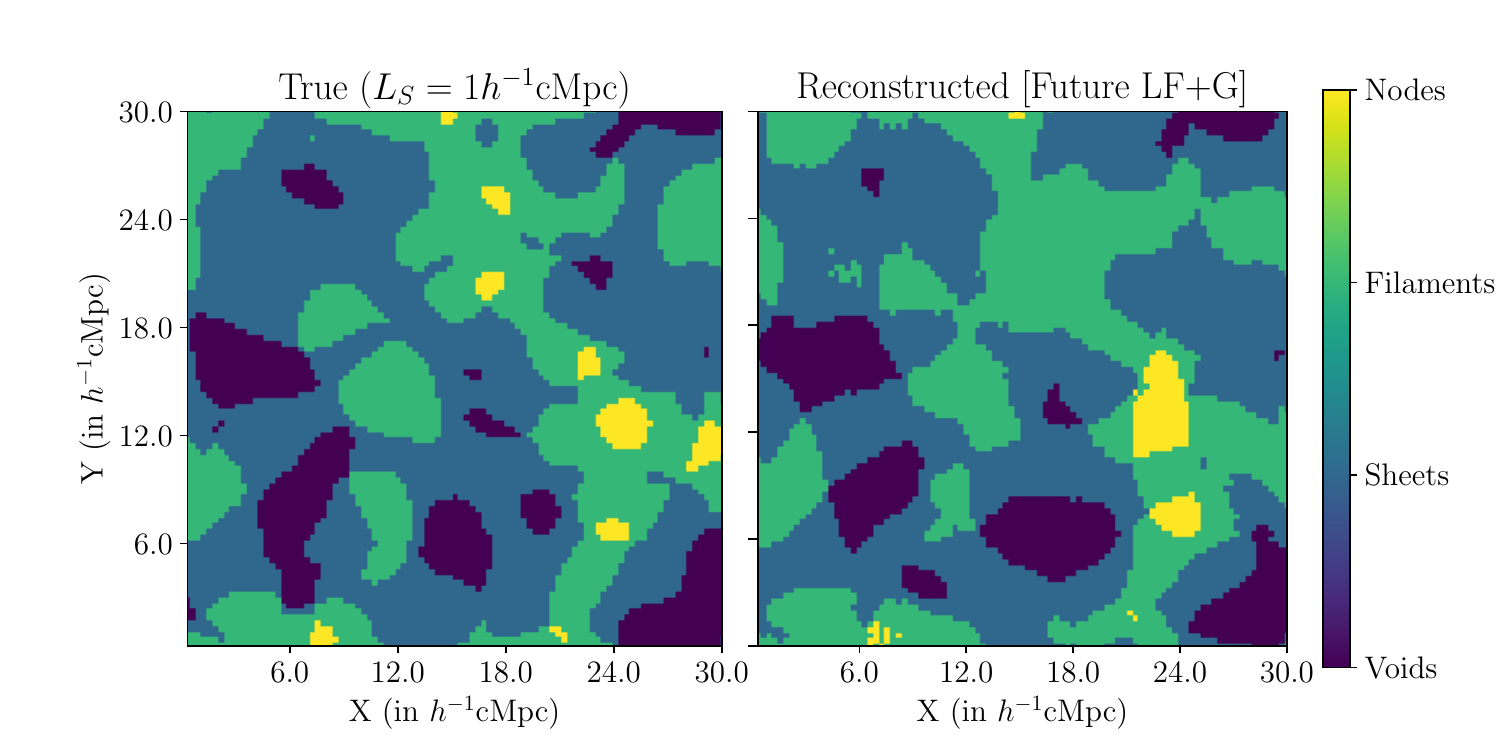}%
            \includegraphics[width=0.4\textwidth, trim={0cm 1.1cm 0cm 0.5cm}, clip]{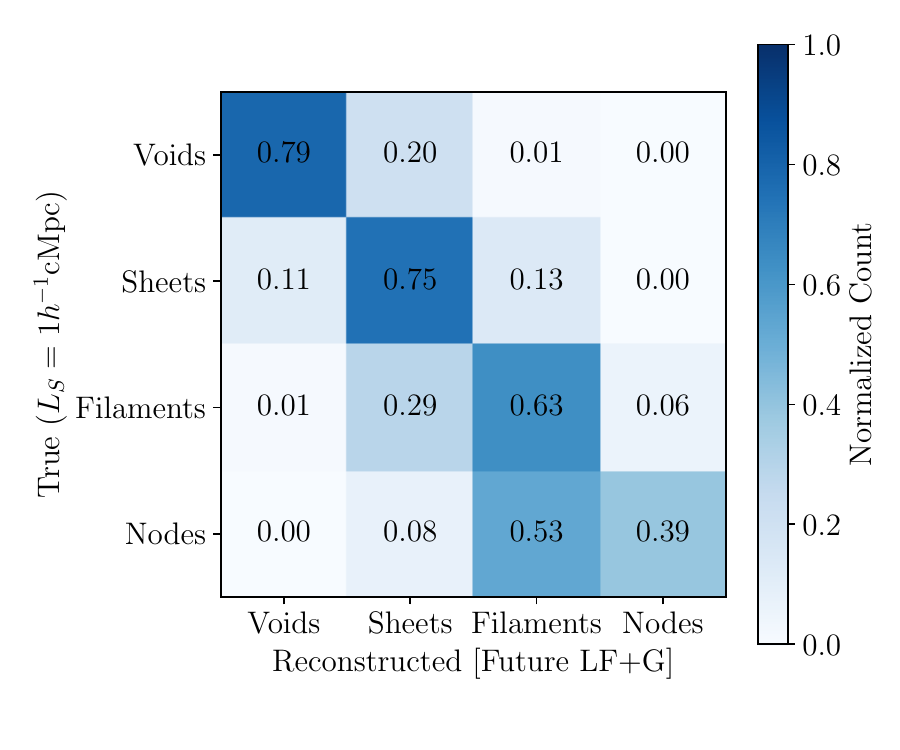}%
        }%
    }

    \caption{
Recovery of cosmic web structures from tomographic reconstructions.
Each row presents a comparison between the true cosmic web classification (left), the reconstructed field (middle), and the corresponding normalized confusion matrix (right) for different survey scenarios and smoothing scales.
The cosmic web environment is defined by the number of eigenvalues of the deformation tensor computed from the gravitational potential above a certain threshold, and is classified into voids, sheets, filaments, and nodes.
\textbf{Top row:} Current survey configuration with mean sightline separation of $2.4\,h^{-1}\mathrm{cMpc}$, with the density field smoothed over $L_S=2\,h^{-1}\mathrm{cMpc}$.
\textbf{Middle and bottom rows:} Future survey configuration with denser sampling ($1\,h^{-1}\mathrm{cMpc}$), shown for $L_S=2\,h^{-1}\mathrm{cMpc}$ and $1\,h^{-1}\mathrm{cMpc}$, respectively.
LF+G denotes reconstructions combining both Ly$\alpha$ forest sightlines and galaxy tracers.
The confusion matrices quantify the fraction of voxels classified as voids, sheets, and filaments in the reconstructed field, conditioned on the true classification.
}

    \label{fig:lss_map}
\end{figure*}

\subsection{Cosmic Web Recovery}\label{sec:web}

An important diagnostic of the reconstructed field is whether it preserves the topological structure of the cosmic web, i.e., the large-scale classification of the cosmic structures into voids, sheets, filaments, and nodes. These environments correspond to distinct patterns of anisotropic collapse and are commonly characterized by the number of eigenvalues above a certain threshold of the deformation tensor derived from the gravitational potential. In this section, we assess how well the {\sc DeepCHART} reconstructions recover these cosmic web environments under both current and future survey conditions.

Following the cosmic web classification method based on the deformation tensor \citep{Hahn2007, ForeroRomero2009}, we define the cosmic web morphology at each voxel by computing the eigenvalues of the deformation tensor:
\begin{equation}
    D_{ij}(\mathbf{x}) = \frac{\partial^2 \Phi(\mathbf{x})}{\partial x_i \partial x_j},
\end{equation}
where \(\Phi(\mathbf{x})\) is the gravitational potential inferred from the overdensity field \(\delta(\mathbf{x}) = \Delta_{\rm DM}(\mathbf{x}) - 1\) via Poisson’s equation in Fourier space. The deformation tensor in Fourier space is computed as:
\begin{equation}
    D_{ij}(\mathbf{k}) = \frac{k_i k_j}{k^2} \tilde{\delta}(\mathbf{k}),
\end{equation}
and subsequently transformed back into real space through an inverse Fourier transform. At each voxel, the local cosmic web environment is determined by counting the number of eigenvalues of \(D_{ij}\) that exceed a chosen threshold \(\lambda_{\mathrm{th}}\). In this work, we adopt \(\lambda_{\mathrm{th}} = 0.05\), a value that roughly reproduces the void volume fraction reported in \citet{Horowitz2019}. Given that our reconstructed and true fields operate in a similar density and resolution regime as theirs, this threshold serves as a consistent and physically motivated choice for identifying cosmic web environments. Under this scheme, regions are classified as voids, sheets, filaments, or nodes depending on whether the number of eigenvalues above the threshold is 0, 1, 2, or 3, respectively.

Figure~\ref{fig:lss_map} displays a comparative view of the cosmic web classification for three different survey scenarios. Each row presents a single voxel-thick 2D slice in the $XY$ plane showing (left) the true cosmic web classification, (middle) the reconstructed classification using {\sc DeepCHART}, and (right) the corresponding normalized confusion matrix that quantifies classification accuracy.
The confusion matrix provides a voxel-wise statistical comparison between the predicted and true structure types: voids, sheets, filaments and nodes. Each element in the matrix represents the fraction of voxels belonging to a given true class (row) that are predicted as a particular class (column). Perfect classification would correspond to a diagonal matrix, while off-diagonal terms indicate misclassification. This framework enables a compact summary of how well the reconstruction preserves the large-scale structural identity of each environment.

The top row in Figure~\ref{fig:lss_map} corresponds to the current survey configuration, characterized by a mean Ly$\alpha$ forest sightline separation of $2.4\,h^{-1}\mathrm{cMpc}$ and a galaxy population consistent with current spectroscopic surveys. The analysis is performed at a smoothing scale of \(L_S = 2\,h^{-1}\mathrm{cMpc}\). The reconstructed field, based on combined LF+G input, visually captures several prominent cosmic web features, including the spatial coherence of some underdense void regions and the connectivity of filamentary structures threading through higher-density environments. The accompanying confusion matrix indicates that voids are most accurately recovered (72\%), followed by sheets (65\%), filaments (50\%), and nodes (32\%). The most significant misclassification patterns include 50\% of true nodes being misclassified as filaments, 39\% of true filaments misclassified as sheets, and 27\% of true voids identified as sheets. These trends are consistent with the overlapping density and morphological characteristics of neighboring cosmic web types. Notably, there is a general tendency for cosmically overdense environments such as nodes and filaments to be misclassified as lower-density structures. This is likely a reflection of the limited dynamic range in overdensities for reconstruction accessible under current survey conditions, where the density contrast in reconstructed fields is restricted (e.g., $\Delta_{\rm DM} \in [0.4, 15]$), leading to partial suppression of high-density features.

The middle row of Figure~\ref{fig:lss_map} illustrates the performance under the future survey configuration, still evaluated at a smoothing scale of \(L_S = 2\,h^{-1}\mathrm{cMpc}\), but incorporating denser Ly$\alpha$ forest sightline sampling and a galaxy population representative of next-generation spectroscopic surveys. The reconstructed map exhibits a notably closer visual correspondence with the true cosmic web morphology, particularly in the enhanced delineation of filamentary ridges and the spatial coherence of underdense voids. These improvements reflect the ability of the deep-learning model to exploit the increased sampling density for more precise structural recovery.
The associated confusion matrix confirms this trend quantitatively. Classification accuracy improves across all web types: voids are now recovered with 81\% accuracy (up from 72\%), sheets increase to 75\% (from 65\%), filaments rise to 63\% (from 50\%) and nodes rise to 43\% (from 32\%). Although the fraction of true nodes misclassified as filaments remains at 50\%, the misclassification of nodes as sheets drops markedly from 18\% to 7\%. This reduction in off-diagonal contamination demonstrates the model’s improved sensitivity to the full range of density environments. In particular, the better recovery of overdense regions such as nodes and filaments likely stems from the expanded dynamic range accessible in the reconstruction under future survey conditions ( $\Delta_{\rm DM} \in [0.25, 40]$), which allows {\sc DeepCHART} to more accurately resolve the nonlinear structures that are partially suppressed in current datasets.

The bottom row of Figure~\ref{fig:lss_map} shows the reconstruction performance at a finer smoothing scale of \(L_S = 1\,h^{-1}\mathrm{cMpc}\), achievable due to the high tracer density expected in future surveys. Visually, the recovered cosmic web reveals sharper structural features, with improved delineation of thin filamentary bridges and smaller-scale voids. Despite the increased spatial complexity introduced at this finer resolution, the reconstruction maintains a high degree of agreement with the true classification.
Quantitatively, the confusion matrix indicates that the accuracy of web-type recovery remains comparable to the coarser smoothing case. Voids are correctly classified 79\% of the time, sheets at 75\%, filaments at 63\% and nodes at 39\%, all within a few percent of their counterparts at \(L_S = 2\,h^{-1}\mathrm{cMpc}\). Misclassifications continue to be dominated by confusion between adjacent morphological classes. But these trends remain consistent with the physical continuity between these structures. These results affirm that the deep-learning framework retains robust performance even at higher resolution, successfully capturing the intricate geometry of the cosmic web without significant degradation in classification fidelity.

Across all cases, the visual maps confirm that the large-scale organization of the cosmic web—void regions, transitional sheets, elongated filaments, and even compact nodes—is broadly preserved in the reconstructed fields. The accompanying confusion matrices provide quantitative support for this, revealing systematic improvements in classification accuracy from current to future survey conditions, and stable recovery even at higher spatial resolution. Notably, {\sc DeepCHART} not only reconstructs isotropic density fluctuations, but also effectively captures the anisotropic features encoded in the deformation tensor, enabling it to recover the intricate morphological structure of the cosmic web from sparse observational tracers. While the classification accuracy for nodes remains somewhat lower compared to other web types, this is likely a consequence of their rarity in volume-limited samples. Increasing the training set to include more realizations where such high-density peaks are well sampled may enhance the model’s ability to reliably recover these compact structures.

\section{Summary and Discussion}
\label{sec:discussion}

We introduce \textsc{DeepCHART} (Deep learning for Cosmological Heterogeneity and Astrophysical Reconstruction via Tomography), a versatile deep learning framework for tomographic reconstruction of underlying astrophysical and cosmological fields from a range of observational tracers. \textsc{DeepCHART} is broadly designed to map various observables, such as absorption features, galaxy distributions, or other large-scale structure tracers, to the underlying physical fields of interest. In this work, we demonstrate its capabilities by reconstructing the three-dimensional dark matter density field at high redshift ($z=2.5$) using Ly$\alpha$ forest absorption and galaxy survey data as inputs. The model is trained on a suite of fully hydrodynamical simulations run with \textsc{GADGET-3}, each covering a comoving volume of $(40,h^{-1}\mathrm{cMpc})^3$ and resolving both dark matter and baryons with $2 \times 512^3$ particles. The training set comprises nine realizations with distinct random seeds, while keeping cosmological and astrophysical parameters fixed to isolate the impact of large-scale structure variance; a tenth realization is reserved for testing. By leveraging a three-dimensional variational autoencoder (VAE) built on a U-Net backbone, our approach enables likelihood-free inference of large-scale structure, capturing both non-linear gravitational evolution and the imprint of baryonic physics present in the simulations. This methodology stands in contrast to traditional analytical and semi-analytical techniques, which are typically constrained by analytical assumptions or computational complexity. Once trained, our model delivers rapid and statistically consistent reconstructions of the cosmic density field from sparse tracers, with each inference requiring only about 0.5 seconds per realization on a GPU. Although we focus here on a fixed thermal and ionization history, the framework is readily extensible to accommodate variations in IGM physics, offering a promising avenue for future explorations and for the further validation of \textsc{DeepCHART} in new physical regimes.

Our experiments span observational regimes reflective of both current and forthcoming Ly$\alpha$ forest surveys, and illustrate the capabilities of \textsc{DeepCHART} in reconstructing the large-scale structure of the high-redshift Universe. For current surveys such as Subaru-PFS and CLAMATO, which achieve typical transverse sightline separations of $2.4\,h^{-1}\mathrm{cMpc}$, we assess reconstruction performance at a smoothing scale of $2\,h^{-1}\mathrm{cMpc}$, a choice that reflects the effective spatial resolution supported by the data. To emulate realistic observational conditions, the simulated Ly$\alpha$ forest spectra are processed through forward modelling that includes instrumental broadening with a Gaussian line-spread function corresponding to a velocity resolution of $\sim120\,\mathrm{km\,s^{-1}}$, and pixel-level Gaussian noise drawn from an SNR distribution ranging from 2 to 10, consistent with expectations for the Subaru-PFS Deep survey. At these survey depths, the model reconstructs the smoothed dark matter density field with reasonable accuracy across a range of cosmic environments. In particular, the reconstruction is most effective in regions with moderate overdensities and underdensities, where the Ly$\alpha$ forest signal retains sensitivity to fluctuations in the underlying matter distribution. The inclusion of galaxy positions alongside Ly$\alpha$ forest sightlines leads to improved recovery in dense environments where the forest saturates and becomes insensitive. Corresponding to matter density contrasts in the interval $0.4 < \Delta_{\mathrm{DM}} < 15$, the voxel-wise correlation between the predicted and true log-density fields reaches $\rho \simeq 0.77$, and the slope of the best-fit relation remains close to unity, indicating that both the spatial structure and amplitude of fluctuations are well preserved. Notably, even when galaxy data are not available at inference time, models trained jointly on both tracers continue to perform better than those trained on forest data alone, suggesting that the network generalises learned mappings from high-density regions. Visual comparisons show that the reconstructed maps recover key morphological features of the cosmic web, including extended voids, broad sheets, and major filamentary structures, although fine-grained contrast is somewhat reduced in the most overdense and underdense regions. 

For future surveys such as WST/IFS and ELT/MOSAIC, which are expected to achieve mean sightline separations as small as $1.0\,h^{-1}\mathrm{cMpc}$, we examine reconstructions at both $2\,h^{-1}\mathrm{cMpc}$ and $1\,h^{-1}\mathrm{cMpc}$ smoothing scales to evaluate the benefits of higher-resolution data. At $2\,h^{-1}\mathrm{cMpc}$ smoothing, the correlation between the reconstructed and true fields increases to $\rho \simeq 0.90$, and remains high ($\rho \simeq 0.86$) even when smoothing is reduced to $1\,h^{-1}\mathrm{cMpc}$, a scale where non-linear structure formation and baryonic processes become increasingly important. In this denser sampling regime, the model recovers the underlying matter field across a broader dynamic range, from deep voids to compact overdensities, covering the interval $0.25 < \Delta_{\mathrm{DM}} < 40$. Visually, the reconstructions display greater contrast and sharpness, with improved resolution of narrow filaments and dense nodes, while preserving the global connectivity of large-scale structures. These improvements reflect the increased constraining power provided by dense tracer sampling and the model’s ability to extract non-linear structure information beyond what is accessible to analytic reconstruction techniques. The use of multiple smoothing scales allows us to systematically assess the limits of spatial information recoverable from realistic survey configurations, and to probe the scales at which physically meaningful structures can be inferred from spectroscopic observations.

To complement the voxel-wise density comparison, we evaluate the one-point probability distribution function (PDF) of the dark matter density contrast to assess how well the reconstructed fields recover the full distribution of density amplitudes. For current survey conditions, the reconstructed PDFs capture the overall non-Gaussian shape of the true field smoothed at $L_S = 2\,h^{-1}\mathrm{cMpc}$ but exhibit suppressed tails, particularly in the deep voids and dense peaks, reflecting the limited sensitivity of the Ly$\alpha$ forest in these regimes. Adding galaxy tracers leads to modest improvements in overdense regions, but residual biases persist near the mean density. In the future survey scenario with denser tracer sampling the reconstructed PDF shows markedly better agreement with the true distribution at both $L_S = 2$ and $1\,h^{-1}\mathrm{cMpc}$ smoothing, with reduced scatter across realizations and improved recovery of the skewness and high-density tail. These results demonstrate that \textsc{DeepCHART} preserves key non-Gaussian features of the field when provided sufficient tracer information, highlighting its ability to capture not only morphology but also the correct amplitude distribution critical for probing non-linear structure formation.

A key validation of any tomographic reconstruction technique lies in its ability to recover the statistical properties of the underlying matter distribution. We evaluate this through a detailed comparison of the spherically averaged power spectra derived from the reconstructed and true density fields. For current survey conditions, we apply a dynamic range cut of $0.4 < \Delta_{\mathrm{DM}} < 15$, while for future surveys with higher tracer densities and improved resolution, the range is broadened to $0.25 < \Delta_{\mathrm{DM}} < 40$. This ensures that power spectrum comparisons are limited to physically meaningful and observationally accessible regions of the field, excluding extreme voids and highly non-linear peaks that are poorly sampled or beyond the dynamic range of the input observables.

Under these conditions, \textsc{DeepCHART} achieves robust recovery of the matter power spectrum across all tracer configurations. For current survey scenarios with $2\,h^{-1}\mathrm{cMpc}$ smoothing, reconstructions using only the Ly$\alpha$ forest exhibit a mild suppression of power at large scales ($k \lesssim 1.5\,h\,\mathrm{cMpc}^{-1}$), likely due to the limited dynamic range and spatial sampling of the forest. The inclusion of galaxy information significantly improves this recovery across all scales, especially in the non-linear regime ($k \gtrsim 1\,h\,\mathrm{cMpc}^{-1}$), where the forest saturates. In this regime, the power is restored to levels closely matching the true clipped field, with the enhancement introduced by galaxies remaining within the statistical uncertainties of the reconstruction. This highlights the complementary role of galaxies in supplementing forest data, particularly for small-scale clustering.

In the case of future surveys with $1.0\,h^{-1}\mathrm{cMpc}$ sightline separations and smoothing on $1\,h^{-1}\mathrm{cMpc}$ scales, the reconstructed field reproduces the true power spectrum at large and small $k$ with high consistency. However, we observe a moderate suppression, up to $\sim40$\%, at intermediate scales around $k \sim 2\,h\,\mathrm{cMpc}^{-1}$, even after clipping. While the origin of this feature is not yet fully understood, it may reflect the combined effects of network-induced smoothing, resolution limitations intrinsic to the training data, and the challenges of reconstructing mildly non-linear structures that span the transition between linear and small-scale clustering. That this suppression does not persist at the highest $k$ suggests that the network can still learn to recover sharp gradients or compact overdensities when adequately sampled, while its performance at intermediate scales may be more sensitive to the interplay between tracer geometry, latent bottleneck constraints, and finite training volume. A more comprehensive investigation of this feature, including training on larger or more diverse volumes, will be necessary to clarify its origin and assess potential mitigation strategies.

Beyond clustering statistics, we assess the ability of \textsc{DeepCHART} to reconstruct the cosmic web: voids, sheets, filaments, and nodes, using a deformation tensor–based classification of the gravitational potential. For current survey conditions ($2.4\,h^{-1}\mathrm{cMpc}$ sightline spacing, $2\,h^{-1}\mathrm{cMpc}$ smoothing), the network recovers voids and sheets with 72\% and 65\% accuracy, while filaments and nodes reach 50\% and 32\%. Misclassifications predominantly occur between adjacent web types, with 50\% of true nodes identified as filaments and 39\% of filaments misclassified as sheets. These trends reflect the limited dynamic range ($\Delta_{\mathrm{DM}} \in [0.4, 15]$) accessible under current conditions, which hampers the recovery of compact overdense features.
Future-like surveys, with $1.0\,h^{-1}\mathrm{cMpc}$ sampling and broader dynamic range ($\Delta_{\mathrm{DM}} \in [0.25, 40]$), yield substantial improvements: accuracy rises to 81\% for voids, 75\% for sheets, 63\% for filaments, and 43\% for nodes. The misclassification of nodes as sheets drops from 18\% to 7\%, and visual maps confirm enhanced coherence of filaments and sharper nodal intersections. At finer $1\,h^{-1}\mathrm{cMpc}$ smoothing, enabled by the denser sampling, performance remains stable with minimal degradation—voids (79\%), sheets (75\%), filaments (63\%), and nodes (39\%)—while the spatial complexity of the web is more clearly delineated.
These results demonstrate that \textsc{DeepCHART} captures not only local density variations but also the anisotropic features of large-scale structure, with improved sensitivity to the cosmic web morphology as tracer density and resolution increase. The comparatively lower accuracy for nodes reflects their rarity and the challenge of reconstructing highly non-linear peaks, motivating future extensions with larger training volumes.

Taken together, our results position DeepCHART as a versatile and scalable tool for field-level inference in cosmology. Its design is inherently modular, permitting the assimilation of diverse observational tracers and the reconstruction of different target fields. While we have focused here on mapping the dark matter distribution from Ly$\alpha$ forest and galaxy data, the underlying methodology is directly applicable to other observables. For instance, the same architecture can be adapted for weak gravitational lensing, enabling direct inference of the 3D matter distribution from projected shear maps. It could also be trained to recover the ionization or temperature fields from high-redshift 21\,cm intensity mapping, or to infer the topology of reionization from sparse Ly$\alpha$ emitter samples. More generally, DeepCHART provides a blueprint for extending deep generative models to a variety of astrophysical problems where the goal is to reconstruct or emulate complex fields from incomplete, noisy, or heterogeneous data.

As future surveys further increase in depth and coverage, and as simulations reach new levels of physical fidelity, the combination of deep learning and synthetic training sets will become increasingly powerful for cosmological inference. DeepCHART opens the door to fast, robust, and flexible reconstructions that will be essential for fully exploiting the information content of upcoming large-scale structure surveys. Further developments, such as integrating uncertainty quantification, conditioning on additional tracers, or leveraging the latent space for parameter inference, will further enhance the utility of this approach for precision cosmology and astrophysics.

\section*{Acknowledgements}
SM acknowledges support from the Italian Ministry of University and Research through PRIN 201278X4FL, PRIN INAF 2019 \textit{"New Light on the Intergalactic Medium"}, and the \textit{Progetti Premiali} funding scheme during part of the period in which this work was carried out. MV is supported by the PD51-INFN INDARK grant. The simulations used in this work were performed using the Ulysses supercomputer at SISSA. This paper is supported by: the Italian Research Center on High Performance Computing, Big Data and Quantum Computing (ICSC), project funded by European Union - NextGenerationEU - and National Recovery and Resilience Plan (NRRP) - Mission 4 Component 2, within the activities of Spoke 3, Astrophysics and Cosmos Observations.  GK acknowledges support from the Kavli Institute Medium Term Visitor programme at the Kavli Institute of Cosmology Cambridge. GK is also partly supported by the Department of Atomic Energy (Government of India) research project with Project Identification Number RTI 4002. The authors acknowledge the computational resources provided by the Department of Theoretical Physics, Tata Institute of Fundamental Research (TIFR). We also acknowledge the use of OpenAI's ChatGPT for support with code implementation, text refinement, and certain analytical calculations. 

\section*{Data Availability}
The data underlying this article will be shared upon reasonable request to the corresponding author. We make \textsc{DeepCHART} publicly available at \url{https://github.com/soumak-maitra/DeepCHART}.



\bibliographystyle{mnras}
\bibliography{main} 




\appendix

\begin{figure*}
    \centering
            \includegraphics[width=0.52\textwidth, trim={1cm 0 0 0.5cm}, clip]{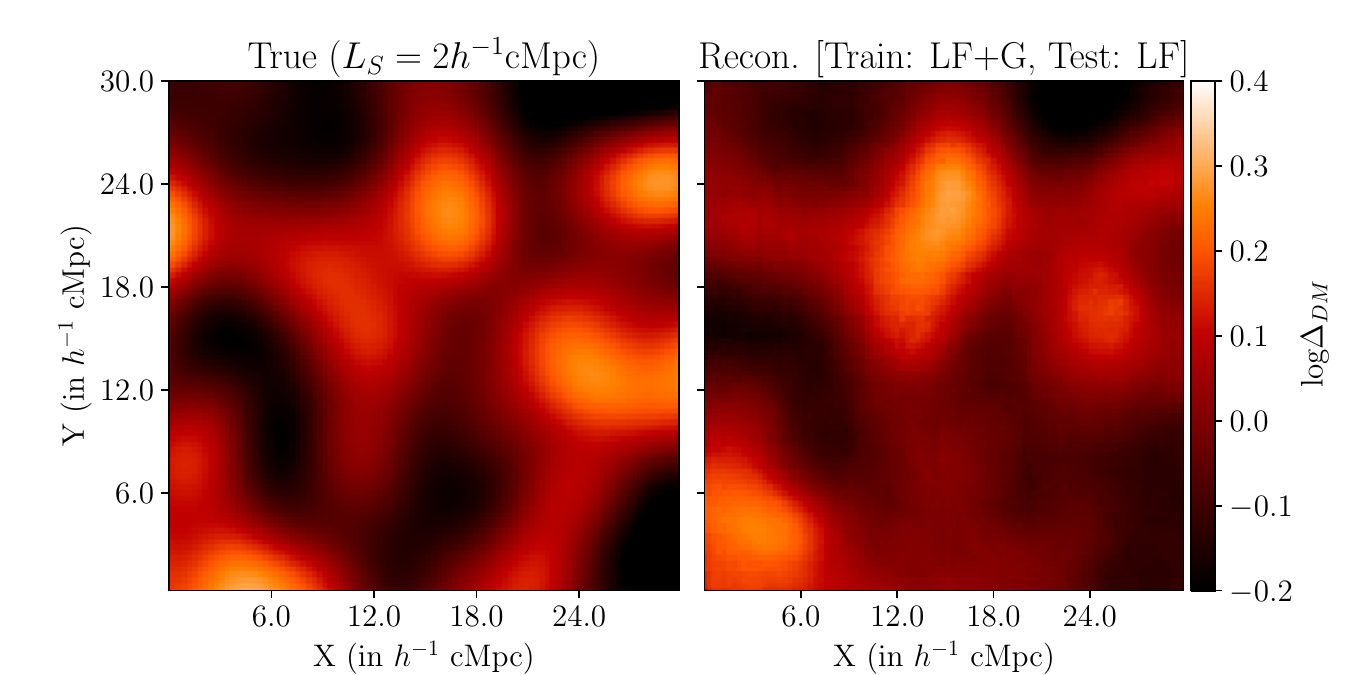}%
            \includegraphics[width=0.48\textwidth, trim={1cm 0 1.8cm 0.5cm}, clip]{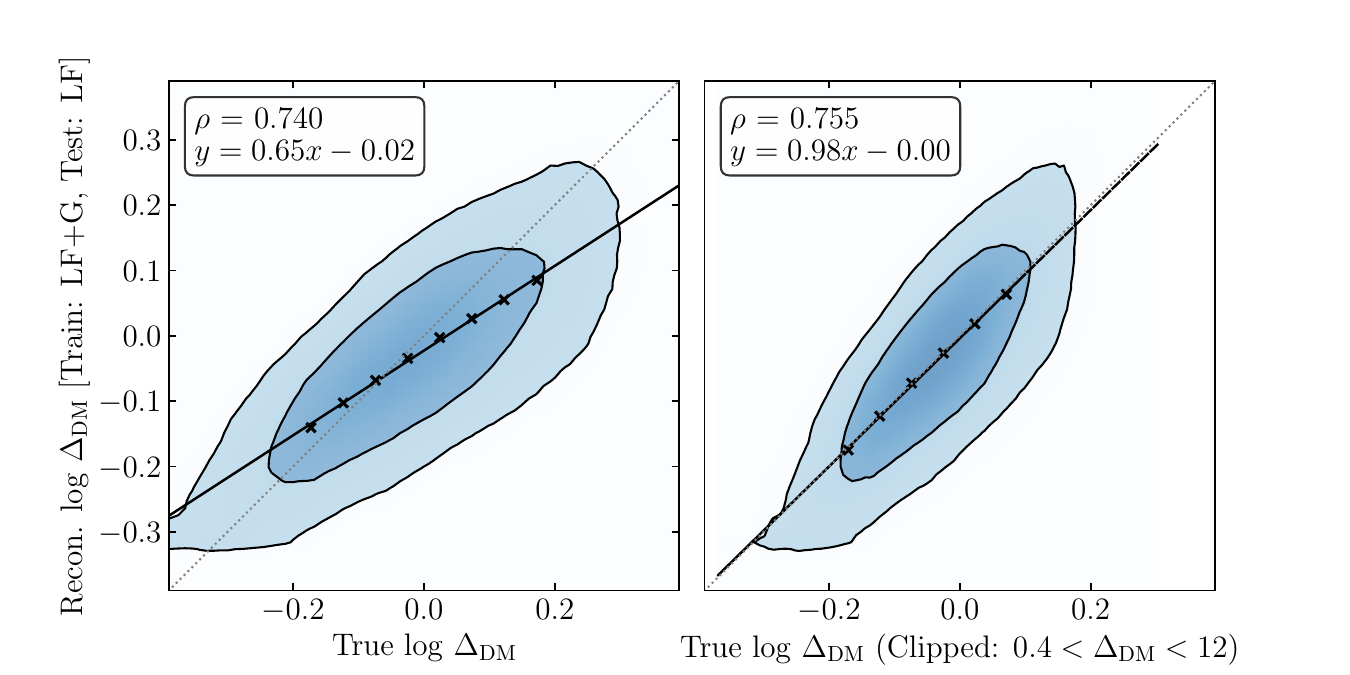}%
        \caption{
LF-limited inference with LF+G-trained model. From left to right: (1) True dark matter field smoothed over \(L_S = 2\,h^{-1}\mathrm{cMpc}\); (2) Reconstruction using a model trained on LF+G input but evaluated using only LF data; (3) Voxel-wise comparison of the reconstructed and true fields; (4) Same comparison restricted to clipped overdensity range \(0.4 < \Delta_{\mathrm{DM}} < 12\). The model retains high fidelity despite the reduced input, and outperforms a model trained solely on LF data.
}
\label{fig:lfonly_infer}
\end{figure*}

\section{\texorpdfstring{Ly$\alpha$ Forest-Only Inference with Ly$\alpha$ Forest and Galaxy-Trained Model}{Lya Forest-Only Inference with Lya Forest and Galaxy-Trained Model}}
\label{A:Cross-Modal}

Here, we explore the scenario where a model trained on both Ly$\alpha$ forest and galaxy (LF+G) data is subsequently used to perform inference using only Ly$\alpha$ forest input. This test serves as an important probe of the model's generalizability when one of the tracers is missing at inference time.
Figure~\ref{fig:lfonly_infer} shows the visual and statistical results of this experiment for the current survey configuration, with the density field smoothed over $L_S = 2\,h^{-1}\mathrm{cMpc}$. The first panel shows the true smoothed density field, while the second panel displays the reconstruction obtained by applying the LF+G-trained model to LF-only input. Interestingly, compared to the model trained and tested solely on LF data (see Figure~\ref{fig:tomography}), the reconstruction quality is noticeably improved: key features of the large-scale structure, such as the dense peaks, are more accurately reproduced, indicating that the model has learned generalized representations of the density field beyond its immediate input modality.

The scatter plots in the third and fourth panels further quantify this improvement. The correlation coefficient rises slightly compared to LF-only training shown in Figure~\ref{fig:contour_accuracy} ($\rho=0.740$ vs.\ $\rho=0.728$ without clipping; $\rho=0.755$ vs.\ $\rho=0.741$ with clipping), and the slope of the best-fit regression line is closer to unity, indicating better amplitude recovery. This improvement is modest but consistent across both the full and clipped dynamic ranges. Notably, the dynamic range over which the clipped comparison remains accurate also increases slightly—from $\Delta_{\rm DM}<10$ in the LF-only case to $\Delta_{\rm DM}<12$ when the model is trained with LF+G input. However, the performance still falls short of reconstructions that use full LF+G information during both training and inference (cf.\ Figure~\ref{fig:lfonly_infer}).

Physically, this result suggests that the joint LF+G training enables the neural network to learn more complete mappings between the input observables and the underlying matter density, capturing correlations that persist even when galaxy information is absent during inference. This is particularly useful in practical scenarios where galaxy observations may be sparse or unavailable in portions of the survey volume. Nonetheless, for optimal performance across the full density range—especially in overdense regions where galaxies are most informative—access to both tracers during inference remains crucial.


\bsp	
\label{lastpage}
\end{document}